\documentclass[%
aps, %
prx, 
twocolumn, 10pt, 
a4paper, %
amsfonts, amssymb, amsmath]{revtex4-1}
\usepackage[sort&compress]{natbib}
\usepackage{newtxtext}
\usepackage[varg]{newtxmath}
\usepackage{ascmac}
\usepackage{graphicx}
\usepackage{bm}
\usepackage{braket}
\usepackage{mathrsfs}
\usepackage[dvipsnames]{xcolor}
\usepackage{hyperref}
\hypersetup{%
	colorlinks=true,%
	linkcolor=blue, %
	citecolor=blue, %
	urlcolor=blue %
}
\newcommand{\br}{\bm{r}}
\newcommand{\bk}{\bm{k}}

\newcommand{\bq}{\bm{q}}
\newcommand{\zi}{i}
\newcommand{\dd}[1]{\mathrm{d} #1\,}
\renewcommand{\Re}{\mathrm{Re}\,}
\renewcommand{\Im}{\mathrm{Im}\,}
\newcommand{\Ord}[1]{\mathcal{O}(#1)}
\newcommand{\subA}{\mathrm{A}}
\newcommand{\subB}{\mathrm{B}}
\newcommand{\Neel}{{N\'{e}el} }
\allowdisplaybreaks[1]
\begin{document}
%
%
%
%
%
\title{Adiabatic and Nonadiabatic Spin-transfer Torques in Antiferromagnets}
\date{\today}
\author{Junji Fujimoto}
\email[Email:]{junji@ucas.ac.cn}
\affiliation{Kavli Institute for Theoretical Sciences, University of Chinese Academy of Sciences, Beijing, 100190, China}

\begin{abstract}
Electron transport in magnetic orders and the magnetic orders dynamics have a mutual dependence, which provides the key mechanisms in spin-dependent phenomena.
Recently, antiferromagnetic orders are focused on as the magnetic order, where current-induced spin-transfer torques, a typical effect of electron transport on the magnetic order, have been debatable mainly because of the lack of an analytic derivation based on quantum field theory.
Here, we construct the microscopic theory of spin-transfer torques on the slowly-varying staggered magnetization in antiferromagnets with weak canting.
In our theory, the electron is captured by bonding/antibonding states, each of which is the eigenstate of the system, doubly degenerates, and spatially spreads to sublattices because of electron hopping.
The spin of the eigenstates depends on the momentum in general, and a nontrivial spin-momentum locking arises for the case with no site inversion symmetry, without considering any spin-orbit couplings.
The spin current of the eigenstates includes an anomalous component proportional to a kind of gauge field defined by derivatives in momentum space and induces the adiabatic spin-transfer torques on the magnetization.
Unexpectedly, we find that one of the nonadiabatic torques has the same form as the adiabatic spin-transfer torque, while the obtained forms for the adiabatic and nonadiabatic spin-transfer torques agree with the phenomenological derivation based on the symmetry consideration.
This finding suggests that the conventional explanation for the spin-transfer torques in antiferromagnets should be changed.
Our microscopic theory provides a fundamental understanding of spin-related physics in antiferromagnets.
\end{abstract}
\maketitle
\section{Introduction}

Manipulation of antiferromagnetic orders by electric means is one of the most important topics~\cite{macdonald2011,jungwirth2016,baltz2018,zelezny2018} because of its applicational potentials, such as producing no stray fields and showing ultrafast dynamics, compared to ferromagnets.
Although the application is an essential driving factor, the phenomena induced by the interplay of the magnetic order with electron transport contains rich physics, which is not fully understood yet.
The antiferromagnetic order is purely quantum mechanical, and the electronic eigenstate coupled with the antiferromagnetic order is no longer \textit{bare} electron, in contrast to the ferromagnetic order.
Here, we call the eigenstate in antiferromagnets the bonding/antibonding state.

The spin-transfer torques in antiferromagnets without spin-orbit couplings have been studied more than for a decade for spin-valve-like structures~\cite{nunez2006,wei2007,urazhdin2007,haney2008,gomonay2010,saidaoui2014,cheng2014} and for slowly-varying antiferromagnetic texutres~\cite{xu2008,swaving2011,hals2011,tveten2013,yamane2016,barker2016,park2020}, as a typical effect of the electron transport on the antiferromagnetic order.
Conventionally, the spin-transfer torque in antiferromagnets has been considered as the spin-transfer torques acting on two coupled ferromagnets, which means that the spin torques are obtained from the summation of the torques on each sublattice magnetization~\cite{xu2008,gomonay2010,park2020}.
However, this explanation is still debatable, mainly because there is no analytic derivation in the adiabatic regime based on an eigenstate picture which clarifies the physics.
Hence, such a fundamental microscopic theory has long been desired.

\begin{figure}[t]
	\centering
	\includegraphics[width=\linewidth]{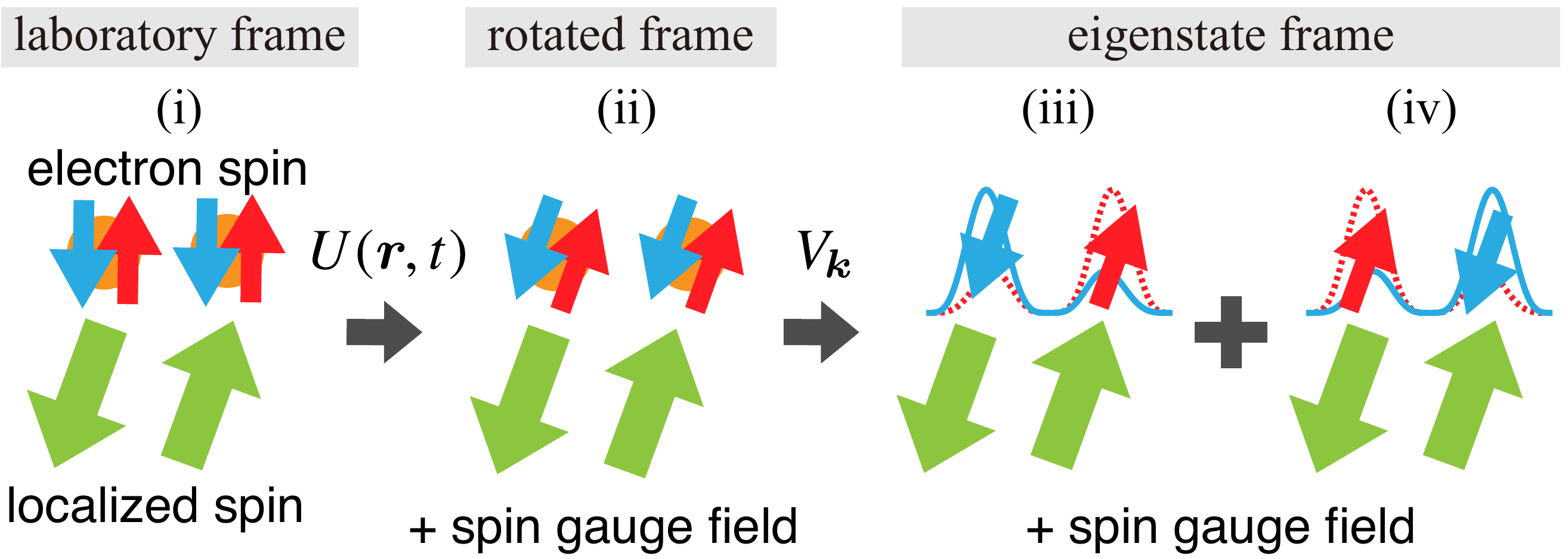}
	\caption{Schematic description of the process for obtaining the effective model by using two unitary transformations $U (\br, t)$ and $V_{\bk}$.
	(i)~The laboratory frame: the quantization axis of the electron spin is independent from space and time.
	(ii)~The rotated frame: by using the unitary transformation $U (\br, t)$, the electron spin is described by the spin coherent state for the \Neel vector.
	In addition, the spatial variation of the \Neel vector is encoded to the spin gauge field in this frame.
	(iii) and (iv)~The eigenstate frame: after the unitary transformation by $V_{\bk}$, the electron is no longer localized at a single sublattice but spreads into both sublattices as (iii) the bonding state and (iv)~the antibonding state, which are the eigenstates of electrons coupled to the \Neel vector.
	}
	\label{fig:unitary-transformation}
\end{figure}

In this paper, we present the microscopic theory based on the eigenstate picture for the spin-transfer torque in slowly-varying staggered antiferromagnets with weak canting.
To derive the model based on the eigenstate picture, we use two unitary transformations as schematically shown in Fig.~\ref{fig:unitary-transformation}.
One is the real space transformation, $U (\br, t)$, in which the quantization axis of the electron spin becomes along the local \Neel vector~\cite{nakane2020}, where we call this frame the rotated frame~(Fig.~\ref{fig:unitary-transformation}~(ii)).
In the rotated frame, the spatial variation of the \Neel vector is encoded to the real space gauge field $\mathcal{A}_{\mathrm{r}, i} = - \zi U^{\dagger} \partial_i U$, which couples to the spin current, and the magnetization due to canting couples to the electron spin.
The other transformation $V_{\bk}$ is the momentum space transformation to diagonalize the Hamiltonian, where we call the frame after this transformation the eigenstate frame.
In the eigenstate frame, we have the Hamiltonian described by the eigenstates, the bonding and antibonding states, each of which doubly degenerates and spreads to sublattices~(Fig.~\ref{fig:unitary-transformation}~(iii) and (iv)).

The perturbation Hamiltonians are also transformed by the unitary matrices $U (\br, t)$ and $V_{\bk}$.
In the eigenstate frame, the electron spin is found to be generally depending on the momentum of the eigenstate, and a nontrivial spin-momentum locking arises for the case with no site inversion symmetry, without considering any spin-orbit couplings.
Note that the site inversion symmetry, or site-centered inversion symmetry, is usually used in the one dimensional~(1D) quantum spin systems and is defined as the invariance under the transformation of site index $i$ changing to $-i$ in the center of the $i = 0$ site for 1D systems~\cite{fuji2015}.
The square lattice is a case with site inversion symmetry, and the honeycomb lattice is a case with no site inversion symmetry~(Fig.~\ref{fig:models}~(b) and (c)).
The point is that the inter-sublattice hopping matrix element is complex in the case with no site inversion symmetry.

The spin current is defined as the current coupled with the real space SU(2) gauge field $\mathcal{A}_{\mathrm{r}, i}$.
In the eigenstate frame, we find that the spin current consists of two components; one is the ordinary spin current given by the combination of the velocity and spin.
The other is an anomalous spin current given by the momentum space gauge field defined by $\mathcal{A}_{\mathrm{k}, i} = - \zi V_{\bk}^{\dagger} \partial_i V_{\bk}$.
A similar momentum space gauge field was discussed by Cheng and Niu~\cite{cheng2012}, but it seems to be different from $\mathcal{A}_{\mathrm{k}, i}$.

We then evaluate the effect of the conduction electron on the dynamics of the antiferromagnets, which is the spin-transfer torque on the order parameters; the torques on the \Neel vector denoted by $\bm{\tau}_n$, and the torque on the magnetization denoted by $\bm{\tau}_m$.
Considering the antiferromagnetically ordered localized spin system, we derive the equation of motion of the order parameters in the presence of the $sd$ exchange coupling to the conduction electron spin.
In the adiabatic regime, the spin torque $\bm{\tau}_m$ is given by the divergence of the anomalous spin current, which is a novel expression, and the other torque $\bm{\tau}_n$ is proportional to the perpendicular component of the conduction electron spin, which is the same form as the spin torque in the ferromagnets.
By evaluating the linear responses of anomalous spin current and spin to the electric field, we obtain the current-induced spin-transfer torques.

As mentioned above, the spin-transfer torque in antiferromagnets has been considered conventionally as the spin-transfer torques acting on two coupled ferromagnets.
For the \textit{adiabatic} spin-transfer torques defined as the spin-transfer torque to which the adiabatic processes only contribute, we confirm that the above explanation is valid.
This agreement is because, in the adiabatic regime, each of the bonding/antibonding states conducts each sublattice.

We also evaluate the \textit{nonadiabatic} spin-transfer torques defined as the torques to which the nonadiabatic processes such as the mixing of the bonding and antibonding states contribute.
We find that one of the nonadiabatic spin-transfer torques $\bm{\tau}_{n}^{\mathrm{na}}$ (the superscript `na' denotes the nonadiabatic) has the same dependence as the adiabatic torque $\bm{\tau}_{n}$, which means that the above conventional explanation does not work for the nonadiabatic torques.
This deviation is understood by the fact that the nonadiabatic processes are characteristic of electrons coupled to the antiferromagnetic order, which are not equivalent to the electrons coupled to two coupled ferromagnetic orders.

We here give a comment on the relation of the adiabatic/nonadiabatic torques to the reactive/dissipative torques.
The reactive and dissipative spin-transfer torques are defined as the even and odd terms under the time-reversal transformation in the equation of motion of the magnetic orders.
In ferromagnets, the reactive torque is equivalent to the adiabatic one, and the dissipative is the same as the nonadiabatic.
However, in antiferromagnets, the correspondences do not realize, since one of the nonadiabatic torques has the same form as the adiabatic torque.

The paper is organized as follows.
In Sec.~\ref{sec:model}, we define the electron system we consider, and Sec.~\ref{sec:effective_model} is devoted to the derivation of the effective Hamiltonian in the eigenstate frame.
In Sec.~\ref{sec:stt}, we present the main topic of the adiabatic and nonadiabatic spin-transfer torques.
Then, Sec.~\ref{sec:conclusion} concludes this paper.

\section{\label{sec:model}Model}
We begin with the tight-binding model coupled to the staggered magnetization slowly-varying spatially through the $sd$-type exchange coupling, in which the Hamiltonian is given by $\mathcal{H}_e = \mathcal{H}_{t} + \mathcal{H}_{sd} + V_{\rm imp}$.
The first term is hopping Hamiltonian given by $\mathcal{H}_t = \sum_{i,j} ( t_{ij} c_{i}^{\dagger} c_{j}^{} + \mathrm{H.c.} ) $ with $t_{ij}$ the hopping integral, which is assumed to be finite only for the nearest-neighbors of the intra-sublattices and of inter-sublattices, where $c_{i}$ is the spinor form of the annihilation operator on the $i$-th site.
The second term is the exchange coupling given by $\mathcal{H}_{sd} = - J_{sd} \sum_{i} \bm{S}_{i} \cdot ( c_{i}^{\dagger} \bm{\sigma} c_{i}^{} )$ with the strength $J_{sd} (>0)$ and the localized spin $\bm{S}_i$ which consists of the staggered magnetization with weak canting.
Here, $\bm{\sigma} = (\sigma^x, \sigma^y, \sigma^z)$ is the Pauli matrices for spin space, and we use $\sigma^0$ as the unit matrix for spin space.
$V_{\mathrm{imp}}$ denotes an impurity potential, which is to be approximated as a nonmagnetic potential acting on the eigenstates for simplicity, and leads to $\bk$-independent lifetime of the bonding/antibonding states.
The impurity effects are out of focus in this paper, and we will discuss them another paper.

\begin{figure*}[bpth]
\centering
\includegraphics[width=\linewidth]{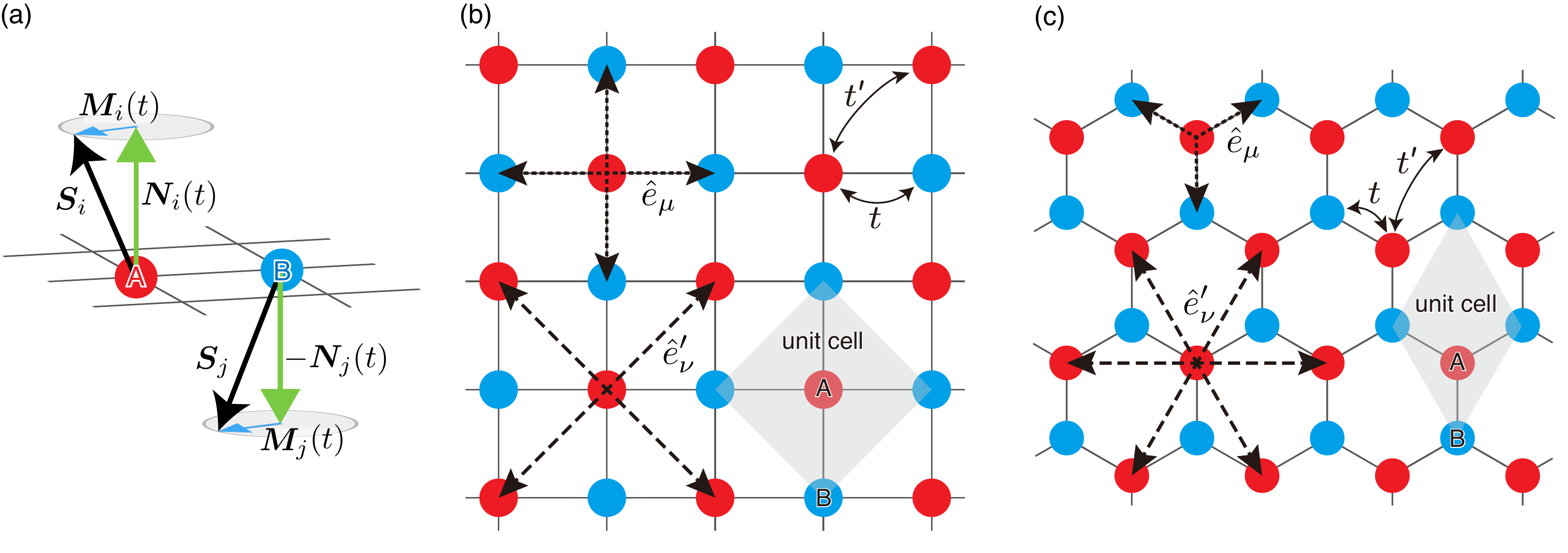}
\caption{\label{fig:models}
(a) The decomposition of the localized spin into the \Neel vector and the magnetization.
From the decomposition, we can define the adiabatic motion of electron for the antiferromagnetic ordered state, since $\bm{N}_i (t) = \bm{N} (\br_i, t)$ and $\bm{M}_i (t) = \bm{M} (\br_i, t)$ are slowly-varying vector, while the vector $\bm{S}_i$ is rapidly changing.
(b) and (c) Schematic figures of (two dimensional) antiferromagnetic models; (b)~the square lattice, (c)~honeycomb lattice.
Then intra-sublattice nearest neighbor vector is described by $\hat{e}'_{\nu}$, and inter-sublattice nearest neighbor vector is denoted by $\hat{e}_{\mu}$.
}
\end{figure*}

In antiferromagnets with the slowly-varying staggered magnetization and with weak canting, the localized spin $\bm{S}_i$ is expressed by using two smooth functions; the \Neel vector $\bm{N} (\br, t)$ and the magnetization $\bm{M} (\br, t)$ as $\bm{S}_i = (-1)^{P_{i}} \bm{N} (\br_i, t) + \bm{M} (\br_i, t)$, (see Fig.~\ref{fig:models} (a)), where $\br_i$ is the position of the $i$-th site, and $P_i$ describes the sign change depending on the sublattice; $P_i = 0$ for $i \in \subA$ and $P_i = 1$ for $i \in \subB$.
For the case of weak canting, the \Neel vector is much larger than the magnetization; $M / N \ll 1$ with $\bm{N} (\br, t) = N \bm{n} (\br, t)$ and $\bm{M} (\br, t) = M \bm{m} (\br, t)$, where $\bm{n}$ and $\bm{m}$ are the unit vectors.
We here assume $N$ and $M$ as well as $S = |\bm{S}_i|$ are constant for time and space, which leads $\bm{n} \cdot \bm{m} = 0$.

\section{\label{sec:effective_model}Effective Hamiltonian}
We now derive the Hamiltonian for the bonding/antibonding states (Fig.~\ref{fig:unitary-transformation}).
Firstly, we take the unitary transformation $U (\br, t)$ such that the electron spin takes the spin coherent state along the \Neel vector; $U^{\dagger} ( \bm{n} \cdot \bm{\sigma} ) U = \sigma^z$ with $c_{i} = U \tilde{c}_i$.
Using the spin gauge field~\cite{korenman1977,bazaliy1998,kohno2007} $\mathcal{A}_{\mathrm{r}, \mu} = - \zi U^{\dagger} (\partial U / \partial r_{\mu}) = A_{\mathrm{r}, \mu}^{\alpha} (\br, t) \sigma^{\alpha} / 2$ with $(r_0, r_1, r_2, r_3) = (t, x, y, z)$ and $\alpha = x, y, z$, we find 
\begin{align}
\mathcal{H}_e
	& = \mathcal{H}_0 + \mathcal{H}_{A} + \mathcal{H}_{f} + \Ord{\mathcal{A}_{\mathrm{r}}^2}
\label{eq:Hamiltonian}
\end{align}
with $\mathcal{H}_{0} = \sum_{\bk} \Psi_{\bk}^{\dagger} H_{\bk} \Psi_{\bk}^{} + V_{\rm imp}$ and
\begin{align}
\mathcal{H}_{A}
	& = \frac{\hbar}{2} \int_{\omega} \sum_{\bq} 
		j_{s, i}^{\alpha} (- \bq) A^{\alpha}_{\mathrm{r}, i} (\bq, \omega)
\label{eq:H_A}
, \\
\mathcal{H}_{f}
	& = - \frac{M}{N} \mathcal{J}_{sd} \int_{\omega} \sum_{\bq}
		s^{\alpha} (- \bq) \bigl( \mathcal{R}^{-1} \bm{m} \bigr)_{\bq, \omega}^{\alpha}
\label{eq:H_f}
,\end{align}
where we introduced $\Psi_{\bk}^{}$ as two sets of spinors of annihilation operators, $H_{\bk}$ describes the unperturbed Hamiltonian given by $ H_{\bk} = T_{\bk} \rho_0 \sigma^0 - (\Re \eta_{\bk}) \rho_1 \sigma^0 - (\Im \eta_{\bk}) \rho_2 \sigma^0 - \mathcal{J}_{sd} \rho_3 \sigma^z $ with $\rho_{\mu}$ being the Pauli matrix for sublattice space for $\mu = 1, 2, 3$ and being the unit matrix for $\mu = 0$.
Here, $T_{\bk}$ and $\eta_{\bk}$ are hoppings of intra- and inter-sublattices, respectively, and we introduced $\mathcal{J}_{sd} = N J_{sd}$.
In this work, we assumed that the electron is described by the inter- and intra-sublattice nearest neighbor hoppings as shown Fig.~\ref{fig:models} (b) and (c), we find $T_{\bk} = - 2 t' \sum_{\nu} \Re [ \exp( - \zi \bk \cdot \hat{e}'_{\nu} ) ]$ and $\eta^{}_{\bk} = t \sum_{\mu} \exp( - \zi \bk \cdot \hat{e}_{\mu} )$, where $t$ and $t'$ are inter- and intra-sublattice nearest neighbor hopping parameters, respectively, and $\hat{e}_{\mu}$ and $\hat{e}'_{\nu}$ are inter- and intra-sublattice nearest neighbor vectors.
The Hamiltonian $\mathcal{H}_A$ represents the couplings of the spin current $j_{s, i}^{\alpha} (\bq)$ to the spin gauge fields, where $\alpha = x, y, z$ is the spin index, and $i = x, y, z$ is spatial index.
The Hamiltonian $\mathcal{H}_{f}$ describes the coupling of the magnetization and the spin in the conventional way of the exchange coupling, where $M / N \ll 1$ assures that we can treat $\mathcal{H}_f$ perturbatively.
Here, $\mathcal{R}$ is the rotational matrix defined by $U^{\dagger} (\br, t) \bm{\sigma} U (\br, t) = \mathcal{R} \bm{\sigma}$.
The frame described by $\Psi_{\bk}$ is called the \textit{rotated} frame~(Fig.~\ref{fig:unitary-transformation}).

Secondly, $H_{\bk}$ is not yet a diagonal matrix, hence necessary to be diagonalized to obtain the effective model.
We diagonalize $H_{\bk}$ so as to hold the equation
\begin{align*}
V_{\bk}^{\dagger} H_{\bk} V_{\bk}^{}
	& = T_{\bk} \rho_0 \sigma^0 - \Delta_{\bk} \rho_3 \sigma^z
\notag \\ & =
	\begin{pmatrix}
		T_{\bk} - \Delta_{\bk}
	&
	&
	&
	\\	
	&	T_{\bk} + \Delta_{\bk}
	&
	&
	\\	
	&
	&	T_{\bk} + \Delta_{\bk}
	&
	\\	
	&
	&
	&	T_{\bk} - \Delta_{\bk}
	\end{pmatrix}
\end{align*}
with $\Delta_{\bk} = (|\eta_{\bk}|^2 + \mathcal{J}_{sd}^2)^{1/2}$.
The explicit form of $V_{\bk}$ is given in Supplemental Material~(SM)~\cite{STTSM}.
From this, we see that the outer $2 \times 2$ matrix is corresponding to the bonding state and the inner $2 \times 2$ matrix is the antibonding state.
The spin and spin current operator in Eqs.~(\ref{eq:H_A}) and (\ref{eq:H_f}) are also transformed by $V_{\bk}$.

Thirdly, we use the projection operator $P_{\pm} = (\rho_0 \sigma^0 \pm \rho_3 \sigma^z)/2$ to the bonding state for $s = +$ and to the antibonding state for $s = -$, which leads $P_{s} (V_{\bk}^{\dagger} \Psi_{\bk}) = \psi_{\bk, s}$, where $\psi_{\bk, s}$ is annihilation operator of the bonding/antibonding state, which can be written by using that of the spin coherent states as
\begin{align}
\psi_{\bk, s}
	& = s \begin{pmatrix}
		\cos (\vartheta_{\bk}/2) c^{\subA}_{\bk, s}
		+ s e^{-\zi \varphi_{\bk}} \sin (\vartheta_{\bk}/2) c^{\subB}_{\bk, s}
	\\[1ex]
		\cos (\vartheta_{\bk}/2) c^{\subB}_{\bk, \bar{s}}
		+ s e^{ \zi \varphi_{\bk}} \sin (\vartheta_{\bk}/2) c^{\subA}_{\bk, \bar{s}}
	,\end{pmatrix}
\label{eq:annihilation_operator-bonding-antibonding_state}
\end{align}
with $\bar{s} = - s$, where $c^{X}_{\bk, \xi}$ is the annihilation operator of sublattice $X = \subA, \subB$ with the spin coherent state ($\xi = +$) and that antiparallel to the state ($\xi = -$), and we introduced
\begin{align}
\sin \vartheta_{\bk} = |\eta_{\bk}| / \Delta_{\bk}
, \qquad \cos \vartheta_{\bk} = \mathcal{J}_{sd} / \Delta_{\bk}
\end{align}
and
\begin{align}
\varphi_{\bk} = \tan^{-1} (\Im \eta_{\bk} / \Re \eta_{\bk})
.\end{align}
Here, $\varphi_{\bk}$ depends on $\Im \eta_{\bk}$, which is only finite in the case with no site inversion symmetry.
We see below that the spin operator of the eigenstates has an essentially different form depending on whether the system has the site inversion symmetry.

We finally obtain the effective model in the adiabatic regime where the mixing of the bonding and antibonding states is negligible, which is given by Eq.~(\ref{eq:Hamiltonian}) with
\begin{align}
\mathcal{H}_{0}
	& = \sum_{s = \pm} \sum_{\bk} \epsilon_{\bk s} \psi_{\bk, s}^{\dagger} \psi_{\bk, s}^{}
		+ V_{\rm imp}
, \quad
\epsilon_{\bk s} = T_{\bk} - s \Delta_{\bk}
\label{eq:H_0}
\end{align}
and Eqs.~(\ref{eq:H_A}) and (\ref{eq:H_f}).
Note that the bonding/antibonding states doubly degenerates, and hence the spinor form can be captured by the another Pauli matrices, which is hereafter denoted by $\tau^{\mu}$ ($\mu = 0, x, y, z$).
We call the frame described by $\psi_{\bk, s}$ the \textit{eigenstate} frame.
We here consider the impurity potential as $V_{\mathrm{imp}} = u_{\mathrm{i}} \sum_{\bk, \bk', s} \rho (\bk' - \bk) \psi_{\bk', s}^{\dagger} \psi_{\bk, s}$ for simplicity, where $u_{\mathrm{i}}$ is the potential strength, and $\rho (\bq)$ is the Fourier component of the impurity density.
The spin in Eq.~(\ref{eq:H_f}) in the eigenstate frame with adiabatic approximation is
\begin{align}
s^{\alpha} (\bq)
	& = \sum_{s = \pm} \sum_{\bk} \psi_{\bk-\frac{\bq}{2}, s}^{\dagger} \left(
			s \tau_{\bk, s}^{\alpha}
		\right) \psi_{\bk+\frac{\bq}{2}, s}^{}
\label{eq:spin}
,\end{align}
where $\bm{\tau}_{\bk, s} = (\tau_{\bk, s}^x, \tau_{\bk, s}^y, \tau_{\bk, s}^z)$ is given as
\begin{align}
\bm{\tau}_{\bk, s}
	& = \begin{pmatrix}
		\sin \vartheta_{\bk} \hat{e}_{\varphi_{\bk}} \cdot \bm{\tau}^{\perp}
	\\	s \sin \vartheta_{\bk} \bigl( \hat{z} \times \hat{e}_{\varphi_{\bk}} \bigr) \cdot \bm{\tau}^{\perp}
	\\	\tau^z
	\end{pmatrix}
\label{eq:spin-vector-of-bonding_state}
\end{align}
with $\bm{\tau}^{\perp} = (\tau^x, \tau^y, 0)$ and $\hat{e}_{\varphi_{\bk}} = (\cos \varphi_{\bk}, \sin \varphi_{\bk}, 0)$.
The spin beyond the adiabatic regime is given in SM~\cite{STTSM}.

Here, we consider a specific configuration, such as the square lattice~(Fig.~\ref{fig:models} (b)), where the inter-sublattice hopping $\eta_{\bk}$ becomes real, $\eta_{\bk} = \eta_{\bk}^*$, hence $\varphi_{\bk} = 0$ and $\hat{e}_{\varphi_{\bk}} = \hat{x}$, we find the spin of the eigenstate as
\begin{align*}
\bm{\tau}_{\bk, s}
	& = \begin{pmatrix}
		\sin \vartheta_{\bk} \tau^x
	\\	s \sin \vartheta_{\bk} \tau^y
	\\	\tau^z
	\end{pmatrix}
.\end{align*}
From this, we find the following three points: (i)~spin direction is independent from the momentum in this case, and the correspondences between the electron picture and the eigenstate picture arises;
\begin{align*}
\sigma^x \leftrightarrow \tau^x
, \quad
\sigma^y \leftrightarrow \tau^y
, \quad
\sigma^z \leftrightarrow \tau^z
,\end{align*}
(ii)~the transverse spin shrinks depending on $\bk$ as $\sin \vartheta_{\bk} = |\eta_{\bk}| / \Delta_{\bk}$, and (iii)~the chirality for the bonding state ($s = +$) is opposite for the antibonding state ($s = -$) since the sign of the $y$ component depends on the bonding or antibonding states.

For the case with no site inversion symmetry, such as the honeycomb lattice~(Fig.~\ref{fig:models} (c)), where the inter-sublattice hopping does not become real, the vector $\hat{e}_{\varphi_{\bk}}$ changes depending on $\bk$, which leads to the momentum-dependent spin direction changing~(see Eq.(\ref{eq:spin-vector-of-bonding_state})).
We emphasise that since we does not consider any spin-orbit couplings, this spin-momentum locking is a nontrivial result.
This spin-momentum locking without any spin-orbit couplings is one of the important findings in this work.
The spin-momentum locking is expected to connect to the nontrivial spin polarization~\cite{hayami2020,yuan2020} and the anomalous Hall effect~\cite{onoda2004} in noncollinear antiferromagnets.

The spin current in Eq.~(\ref{eq:H_A}) is given as $\bm{j}_{\mathrm{s}, i} = (j_{\mathrm{s}, i}^x, j_{\mathrm{s}, i}^y, j_{\mathrm{s}, i}^z)$ with $\bm{j}_{\mathrm{s}, i} (\bq) = \bm{j}_{\mathrm{s0}, i} (\bq) + \bm{j}_{\mathrm{sA}, i} (\bq)$, where $\bm{j}_{\mathrm{s0}, i} (\bq)$ is the ordinary spin current given by
\begin{align}
\bm{j}_{\mathrm{s0}, i} (\bq)
	& = \sum_{\bk, s} \psi_{\bk-\frac{\bq}{2}, s}^{\dagger} \left(
		\frac{s}{\hbar} \frac{\partial \epsilon_{\bk s}}{\partial k_i} \bm{\tau}_{\bk, s}
	\right) \psi_{\bk+\frac{\bq}{2}, s}
\label{eq:j_s0}
,\end{align}
where the spin operator is given by Eq.~(\ref{eq:spin-vector-of-bonding_state}), and $\bm{j}_{\mathrm{sA}, i} (\bq)$ is an anomalous spin current given by
\begin{align}
\bm{j}_{\mathrm{sA}, i} (\bq)
	& = \frac{\mathcal{J}_{sd}}{\hbar} \sum_{\bk, s} \psi_{\bk-\frac{\bq}{2}, s}^{\dagger}
		\begin{pmatrix}
			(\hat{z} \times \bm{A}_{\mathrm{k}, i}^{\perp}) \cdot \bm{\tau}^{\perp}
		\\	- s \bm{A}_{\mathrm{k}, i}^{\perp} \cdot \bm{\tau}^{\perp}
		\\	0
		\end{pmatrix}
	\psi_{\bk+\frac{\bq}{2}, s}
\label{eq:j_sA}
.\end{align}
Here, $A^{\alpha}_{\mathrm{k}, i}$ in Eq.~(\ref{eq:j_sA}) is a kind of gauge filed defined by derivatives in momentum space $\mathcal{A}_{\mathrm{k}, i} = - \zi V^{\dagger}_{\bk} (\partial V_{\bk} / \partial k_i) = (A^z_{\mathrm{k}, i} \rho_3 \sigma^0 + \bm{A}^{\perp}_{\mathrm{k}, i} \cdot \bm{\rho} \sigma^z)/2$ with
\begin{subequations}
\begin{align}
\bm{A}^{\perp}_{\mathrm{k}, i} (\bk)
	& = \frac{\partial \vartheta_{\bk}}{\partial k_i} \hat{z} \times \hat{e}_{\varphi_{\bk}}
		- \frac{\partial \varphi_{\bk}}{\partial k_i} \sin \vartheta_{\bk} \, \hat{e}_{\varphi_{\bk}}
\label{eq:Ak^perp}
, \\
A^z_{\mathrm{k}, i} (\bk)
	& = \frac{\partial \varphi_{\bk}}{\partial k_i} ( 1 - \cos \vartheta_{\bk} )
\label{eq:Ak^z}
.\end{align}
\end{subequations}
The expression of the anomalous spin current is also one of the important points of this work, since this spin current contributes to the spin torque as shown below in Eq.~(\ref{eq:tau_m}).
Note that $A^z_{\mathrm{k}, i} (\bk)$ is the Berry phase in the momentum space and is only finite in the case with no site inversion symmetry.

\section{\label{sec:stt}Spin-transfer torques}
Here, we evaluate the spin-transfer torques on the \Neel vector and magnetization.
First, we derive the equation of motion for the two order parameters in the presence of the $sd$ exchange coupling.
The part related to the localized spin in the Lagrangian is given as $\mathcal{L}_s = \mathcal{L}_B - \mathcal{H}_s - \mathcal{H}_{sd}$, where the first term in the right hand side is calculated in the continuum limit~\cite{sachdev2011} as
$\mathcal{L}_B = - \hbar \int (\dd{\br}/V) [ \bm{M} \cdot \left( \bm{n} \times (\partial \bm{n}/ \partial t) \right) ]$, where $V$ is the volume of the system, and the second term $\mathcal{H}_s$ is the Hamiltonian of the localized spin.
The $sd$ exchange coupling in the continuum limit is rewritten as $\mathcal{H}_{sd} = \int (\dd{\br} / V) [ - N J \bm{n} \cdot \tilde{\bm{\Sigma}} - J \bm{M} \cdot \tilde{\bm{s}} ]$, where $\tilde{\bm{\Sigma}}$ and $\tilde{\bm{s}}$ are staggered electron spin and ferromagnetic electron spin in the laboratory frame, respectively.

The Eular-Lagrange equation is calculated as
\begin{subequations}
\begin{align}
M \dot{\bm{m}}
	& = \bm{n} \times \frac{1}{\hbar} \frac{\delta \mathcal{H}_s}{\delta \bm{n}}
		+ \bm{\tau}_m
		+ \bm{n} \times \frac{1}{\hbar} \frac{\delta \mathcal{W}}{ \delta \dot{\bm{n}} }
, \\
\dot{\bm{n}}
	& = \bm{n} \times \frac{1}{\hbar M} \frac{\delta \mathcal{H}_s}{\delta \bm{m}}
		+ \bm{\tau}_n
		+ \bm{n} \times \frac{1}{\hbar M} \frac{\delta \mathcal{W}}{ \delta \dot{\bm{m}} }
,\end{align}\label{eq:EOM}%
\end{subequations}
where $\mathcal{W}$ is the phenomenologically introduced damping function.
We find that the $sd$ exchange coupling induces the following spin torques in the rotated frame, $\mathcal{R}^{-1} \bm{\tau}_m = (J N / \hbar) (\hat{z} \times \langle \bm{\Sigma} \rangle_{\mathrm{neq}})$ and $\mathcal{R}^{-1} \bm{\tau}_n = (J / \hbar) (\hat{z} \times \langle \bm{s} \rangle_{\mathrm{neq}})$, which are obtained as
\begin{subequations}
\begin{align}
(\mathcal{R}^{-1} \bm{\tau}_{m})_{\bq}
	& = \frac{\zi q_j}{2} \langle \bm{j}_{\mathrm{sA}, j} (\bq, t) \rangle_{\mathrm{neq}}
\label{eq:tau_m}
, \\
(\mathcal{R}^{-1} \bm{\tau}_{n})_{\bq}
	& = \frac{J_{sd}}{\hbar} \hat{z} \times \langle \bm{s} (\bq, t) \rangle_{\mathrm{neq}}
\label{eq:tau_n}
\end{align}\label{eq:tau}%
\end{subequations}
in the adiabatic regime, where $\bm{j}_{\mathrm{sA}, j} (\bq)$ is the anomalous spin current of the eigenstate given by Eq.~(\ref{eq:j_sA}), and the spin $\bm{s} (\bq)$ is given by Eq.~(\ref{eq:spin}).
Here, $\langle \,\cdots \rangle_{\mathrm{neq}}$ means the statistical average in nonequilibrium.
(See SM~\cite{STTSM} for the derivation of Eq.~(\ref{eq:tau_m}).)
Equation~(\ref{eq:tau_m}) suggests that the spin torque on the magnetization $\bm{m}$ is given by the divergence of the anomalous spin current. 
Equation~(\ref{eq:tau_n}) is the same form as the spin torque in ferromagnets.
These two expressions~(\ref{eq:tau}) are one of the important findings in this work.
Note that the spin torque $\bm{\tau}_{m}$ is already the first order of $\bq$ and has no zeroth order terms with respect to $\bq$.

Then, we evaluate the spin torques~(\ref{eq:tau}) in the presence of the electric field.
Following the linear response theory, we have $\langle j_{\mathrm{sA}, i}^{\alpha} (\bq, \omega) \rangle_{\mathrm{neq}} = X_{i j}^{\alpha} (\bq, \omega) A_{j}^{\mathrm{em}} (\omega)$ and $\langle s^{\alpha} (\bq, \omega) \rangle_{\mathrm{neq}} = Y_{j}^{\alpha} (\bq, \omega) A_{j}^{\mathrm{em}} (\omega)$, where $A_{j}^{\mathrm{em}} (\omega)$ is the vector potential, and the linear response coefficients $X_{i j}^{\alpha} (\bq, \omega)$ and $Y_{i}^{\alpha} (\bq, \omega)$ are obtained from the corresponding Matsubara functions $\mathscr{X}_{i j}^{\alpha} (\bq, \zi \omega_{\lambda}) = - e \langle\!\langle j_{\mathrm{sA}, i}^{\alpha} (\bq), j_{j} (0) \rangle\!\rangle$ and $\mathscr{Y}_{j}^{\alpha} (\bq, \zi \omega_{\lambda}) = - e \langle\!\langle s^{\alpha} (\bq), j_{j} (0) \rangle\!\rangle$ with the canonical correlation $\langle\!\langle A, B \rangle\!\rangle = V^{-1} \int_0^{\beta} \dd{\tau} e^{\zi \omega_{\lambda} \tau} \langle \mathrm{T}_{\tau} A (\tau), B (0) \rangle$ by taking the analytic continuation $\zi \omega_{\lambda} \to \hbar \omega + \zi 0$.
Here, $j_j (\bq)$ is the electric current and the explicit form in the eigenstate frame is given in SM~\cite{STTSM}, and $\beta = 1 / k_{\mathrm{B}} T$ is the inverse temperature.
Rewriting the Matsubara functions in terms of the thermal Green function, we expand the Green function up to the first order of the Hamiltonian $\mathcal{H}_f$ for the coefficient $\mathscr{X}_{i j}^{\alpha} (\bq, \zi \omega_{\lambda})$.
For the coefficient $\mathscr{Y}_{j}^{\alpha} (\bq, \zi \omega_{\lambda})$, we expand the first order of the spin gauge field $A_{\mathrm{r}, i}$ as in the calculation of the spin-transfer torques in ferromagnets~\cite{kohno2007,tatara2008,fujimoto2019a}.
Note that the terms proportional to the spin gauge field in $\mathscr{X}_{i j}^{\alpha} (\bq, \zi \omega_{\lambda})$ become the spin torques in the second order of $\bq$ since the spin gauge field is the first order of $\bq$ and Eq.~(\ref{eq:tau_m}) is already the first order of $\bq$, so that we neglect the terms.

After some straightforward calculations~\cite{STTSM}, the resultant expressions for the adiabatic spin-transfer torques in the laboratory frame are obtained as
\begin{subequations}
\begin{align}
\bm{\tau}_{m} (\br, t)
	& = \frac{M}{N} \left( \frac{\mathcal{P}_m}{e} \bm{j}_{\mathrm{c}} \cdot \bm{\nabla} \right) \bm{m} (\br, t)
\label{eq:tau_ad_m}
, \\
\bm{\tau}_{n} (\br, t)
	& = \frac{1}{N} \left( \frac{\mathcal{P}_n}{e} \bm{j}_{\mathrm{c}} \cdot \bm{\nabla} \right) \bm{n} (\br, t)
\label{eq:tau_ad_n}
,\end{align}\label{eq:results}%
\end{subequations}
where $- e$ is the elementary charge, $\bm{j}_{\mathrm{c}}$ is the charge current, and $\mathcal{P}_m$ and $\mathcal{P}_n$ are nondimensional coefficients, which are given in SM~\cite{STTSM}.
Equations~(\ref{eq:results}) are the first main results of this work.
This result does not depend on the lattice symmetry.

The obtained spin-transfer torques $\bm{\tau}_m$ and $\bm{\tau}_n$ are similar forms to that in the ferromagnets $\propto (\bm{j}_{\mathrm{s}} \cdot \bm{\nabla}) \bm{m}'$, where $\bm{j}_{\mathrm{s}}$ is the spin current and $\bm{m}'$ is the magnetization in the ferromagnet.
Especially, Eq.~(\ref{eq:tau_ad_n}) is the same form as Eq.~(20) in Ref.~\cite{park2020}, which is obtained by considering the antiferromagnet as the two coupled ferromagnets. 
This agreement suggests that the conventional explanation is valid for the adiabatic spin-transfer torque.
It is mainly because each of the doubly-degenerated bonding/antibonding state conducts each sublattice in the adiabatic regime.
Note that there is one difference from the case of ferromagnets; in antiferromagnets, $\mathcal{P}_m \bm{j}_{\mathrm{c}}$ and $\mathcal{P}_n \bm{j}_{\mathrm{c}}$ are not equivalent to the spin current.
In ferromagnets, the charge current accompanies with spin polarization, hence charge current can be rewritten as $\bm{j}_{\mathrm{s}} = \mathcal{P} \bm{j}_{\mathrm{c}}$ by means of the spin polarization $\mathcal{P}$, while the charge current in antiferromagnets does not accompany with spin polarization.
The eigenstate is doubly degenerated as Eq.~(\ref{eq:annihilation_operator-bonding-antibonding_state}), and both of two degenerated sates contribute additively to the spin-transfer torques, so that charge current induces the spin torques.

In order to treat a nonadiabatic contribution to the spin torques, we need to consider nonadiabatic processes, which are transitions from the bonding state to the antibonding state and vice versa.
Note that the nonadiabatic torques are not obtained from Eqs.~(\ref{eq:tau}) even with the spin relaxation mechanism.
The derivation and calculation of the nonadiabatic spin-transfer torques are given in SM~\cite{STTSM}, and we find
\begin{subequations}
\begin{align}
\bm{\tau}_m^{\mathrm{na}} (\br, t)
	& = \zi \omega \tau_{sd} \bm{n} \times \left( \frac{ \mathcal{P}_m^{\mathrm{na}} }{e} \bm{j}_{\mathrm{c}} \cdot \bm{\nabla} \right) \bm{n}
\label{eq:tau_na_m}
, \\
\bm{\tau}_n^{\mathrm{na}} (\br, t)
	& = - \frac{1}{N} \left( \frac{ \mathcal{P}_n^{\mathrm{na}} }{e} \bm{j}_{\mathrm{c}} \cdot \bm{\nabla} \right) \bm{n}
\label{eq:tau_na_n}
,\end{align}\label{eq:tau_na}%
\end{subequations}
where $\tau_{sd} = \hbar / 2 \mathcal{J}_{sd}$, and $\mathcal{P}^{\mathrm{na}}_m$ and $\mathcal{P}^{\mathrm{na}}_n$ are nondimensional coefficients given in SM~\cite{STTSM}.
Here, $\omega$ is the frequency of the external electric field $\bm{E} = \bm{E}_0 e^{-\zi \omega t}$ and the charge current is given by $\bm{j}_{\mathrm{c}} = \sigma_{\mathrm{c}} \bm{E}$, where $\sigma_{\mathrm{c}}$ is the conductivity.
Equations~(\ref{eq:tau_na}) are the second main results of this work.
Note that $\mathcal{P}^{\mathrm{na}}_m = 0$ and $\mathcal{P}^{\mathrm{na}}_n = 0$ for the case where the intra-sublattice hopping $T_{\bk}$ is zero.
In contrast, for the case of $\Delta_{\bk} \simeq \mathcal{J}_{sd}$, we find $\mathcal{P}_m^{\mathrm{na}} = \mathcal{P}_n^{\mathrm{na}} \simeq 1$ (see SM~\cite{STTSM}).

In the previous work~\cite{fujimoto2019a}, we showed that the nonadiabatic spin-transfer torque can be induced by the \textit{alternating} current in ferromagnets.
Equation~(\ref{eq:tau_na_m}) indicates that the \textit{alternating} current induces the nonadiabatic torque in antiferromagnets and the \textit{direct} current ($\omega = 0$) does not give rise to the nonadiabatic torque on $\bm{m}$ for the case of the simple impurity potential we consider.
By analogy of the case in ferromagnets~\cite{fujimoto2019a}, $\zi \omega \tau_{sd}$ could be replaced by $\zi \omega \tau_{sd} + \zeta_{\mathrm{s}}$, where $\zeta_{\mathrm{s}}$ is the spin relaxation rate, and the direct current induces the nonadiabatic spin-transfer torque in that case.

Although the nonadiabatic spin-transfer torque~(\ref{eq:tau_na_m}) is already expected from the symmetry consideration~\cite{hals2011}, the torque~(\ref{eq:tau_na_n}) is unexpectedly obtained.
Equation~(\ref{eq:tau_na_n}) suggests that the nonadiabatic processes also contribute to the ordinary spin-transfer torque~(\ref{eq:tau_ad_n}), which is essentially different from ferromagnets, where the nonadiabatic process only contributes to the nonadiabatic spin-transfer torques.
The result~(\ref{eq:tau_na_n}) indicates that the conventional explanation for the spin-transfer torque does not work for the nonadiabatic torque.
Furthermore, the correspondences between adiabatic (nonadiabatic) torque and reactive (dissipative) torque, which is valid in ferromagnets, are no longer realized due to Eq.~(\ref{eq:tau_na_n}).

For further discussion, we need to solve the equations~(\ref{eq:EOM}) for a specific configuration, such as a domain wall~\cite{papanicolaou1995,jaramillo2007,okuno2019}, and the analysis is one of the future works.

\section{\label{sec:conclusion}Conclusion}
In conclusion, we have constructed the microscopic theory of the spin-transfer torques on the slowly-varying staggered magnetization in antiferromagnets with weak canting.
The effective model is obtained by using the two unitary transformations; one is the real space unitary transformation in which the electron spin is to be along the \Neel vector, and the other is the momentum space unitary transformation in which the unperturbed Hamiltonian is to be diagonalized.
By these transformations, we have two kinds of gauge fields; one is the spin gauge field which is well-known in ferromagnetic spintronics, and the other is the momentum space gauge field, which is an extension of the momentum space Berry phase.
The spin operator of the eigenstates depends on the momentum in general, and a nontrivial spin-momentum locking arises in the case with no site inversion symmetry.
The spin current operator of the eigenstates has two components; one is the ordinary spin current operator and the other is the anomalous spin current which is proportional to the momentum space gauge field.
The divergence of the anomalous spin current induces the spin torque on the magnetization in the adiabatic regime, and the spin torque on the \Neel vector is given as the same form as the spin torque in ferromagnets.
The obtained forms for the adiabatic and nonadiabatic spin-transfer torques agree with the phenomenological derivation based on the symmetry consideration.
For the adiabatic torques are understood by the conventional explanation for the spin-transfer torques, but the explanation fails for the nonadiabatic spin-transfer torque.
Our microscopic theory provides a fundamental understanding of spin-related physics in antiferromagnets, which paves the way for developing the antiferromagnetic spintronics.

\begin{acknowledgments}
The author would like to thank G.~Tatara and Y.~Yamane for valuable advices in the early stage of this work.
The author also thank M.~Matsuo, A.~Shitade, C.~Akosa, S.C.~Furuya, and Y. Ominato for their stimulating comments and suggestions.
This work is partially supported by the Priority Program of Chinese Academy of Sciences, Grant No. XDB28000000.
\end{acknowledgments}

\bibliography{reference}

\begin{thebibliography}{34}%
\makeatletter
\providecommand \@ifxundefined [1]{%
 \@ifx{#1\undefined}
}%
\providecommand \@ifnum [1]{%
 \ifnum #1\expandafter \@firstoftwo
 \else \expandafter \@secondoftwo
 \fi
}%
\providecommand \@ifx [1]{%
 \ifx #1\expandafter \@firstoftwo
 \else \expandafter \@secondoftwo
 \fi
}%
\providecommand \natexlab [1]{#1}%
\providecommand \enquote  [1]{``#1''}%
\providecommand \bibnamefont  [1]{#1}%
\providecommand \bibfnamefont [1]{#1}%
\providecommand \citenamefont [1]{#1}%
\providecommand \href@noop [0]{\@secondoftwo}%
\providecommand \href [0]{\begingroup \@sanitize@url \@href}%
\providecommand \@href[1]{\@@startlink{#1}\@@href}%
\providecommand \@@href[1]{\endgroup#1\@@endlink}%
\providecommand \@sanitize@url [0]{\catcode `\\12\catcode `\$12\catcode
  `\&12\catcode `\#12\catcode `\^12\catcode `\_12\catcode `\%12\relax}%
\providecommand \@@startlink[1]{}%
\providecommand \@@endlink[0]{}%
\providecommand \url  [0]{\begingroup\@sanitize@url \@url }%
\providecommand \@url [1]{\endgroup\@href {#1}{\urlprefix }}%
\providecommand \urlprefix  [0]{URL }%
\providecommand \Eprint [0]{\href }%
\providecommand \doibase [0]{http://dx.doi.org/}%
\providecommand \selectlanguage [0]{\@gobble}%
\providecommand \bibinfo  [0]{\@secondoftwo}%
\providecommand \bibfield  [0]{\@secondoftwo}%
\providecommand \translation [1]{[#1]}%
\providecommand \BibitemOpen [0]{}%
\providecommand \bibitemStop [0]{}%
\providecommand \bibitemNoStop [0]{.\EOS\space}%
\providecommand \EOS [0]{\spacefactor3000\relax}%
\providecommand \BibitemShut  [1]{\csname bibitem#1\endcsname}%
\let\auto@bib@innerbib\@empty
\bibitem [{\citenamefont {MacDonald}\ and\ \citenamefont
  {Tsoi}(2011)}]{macdonald2011}%
  \BibitemOpen
  \bibfield  {author} {\bibinfo {author} {\bibfnamefont {A.~H.}\ \bibnamefont
  {MacDonald}}\ and\ \bibinfo {author} {\bibfnamefont {M.}~\bibnamefont
  {Tsoi}},\ }\href {\doibase 10.1098/rsta.2011.0014} {\bibfield  {journal}
  {\bibinfo  {journal} {Phil. Trans. R. Soc. A}\ }\textbf {\bibinfo {volume}
  {369}},\ \bibinfo {pages} {3098} (\bibinfo {year} {2011})}\BibitemShut
  {NoStop}%
\bibitem [{\citenamefont {Jungwirth}\ \emph {et~al.}(2016)\citenamefont
  {Jungwirth}, \citenamefont {Marti}, \citenamefont {Wadley},\ and\
  \citenamefont {Wunderlich}}]{jungwirth2016}%
  \BibitemOpen
  \bibfield  {author} {\bibinfo {author} {\bibfnamefont {T.}~\bibnamefont
  {Jungwirth}}, \bibinfo {author} {\bibfnamefont {X.}~\bibnamefont {Marti}},
  \bibinfo {author} {\bibfnamefont {P.}~\bibnamefont {Wadley}}, \ and\ \bibinfo
  {author} {\bibfnamefont {J.}~\bibnamefont {Wunderlich}},\ }\href {\doibase
  10.1038/nnano.2016.18} {\bibfield  {journal} {\bibinfo  {journal} {Nat.
  Nanotechnol.}\ }\textbf {\bibinfo {volume} {11}},\ \bibinfo {pages} {231}
  (\bibinfo {year} {2016})}\BibitemShut {NoStop}%
\bibitem [{\citenamefont {Baltz}\ \emph {et~al.}(2018)\citenamefont {Baltz},
  \citenamefont {Manchon}, \citenamefont {Tsoi}, \citenamefont {Moriyama},
  \citenamefont {Ono},\ and\ \citenamefont {Tserkovnyak}}]{baltz2018}%
  \BibitemOpen
  \bibfield  {author} {\bibinfo {author} {\bibfnamefont {V.}~\bibnamefont
  {Baltz}}, \bibinfo {author} {\bibfnamefont {A.}~\bibnamefont {Manchon}},
  \bibinfo {author} {\bibfnamefont {M.}~\bibnamefont {Tsoi}}, \bibinfo {author}
  {\bibfnamefont {T.}~\bibnamefont {Moriyama}}, \bibinfo {author}
  {\bibfnamefont {T.}~\bibnamefont {Ono}}, \ and\ \bibinfo {author}
  {\bibfnamefont {Y.}~\bibnamefont {Tserkovnyak}},\ }\href {\doibase
  10.1103/RevModPhys.90.015005} {\bibfield  {journal} {\bibinfo  {journal}
  {Rev. Mod. Phys.}\ }\textbf {\bibinfo {volume} {90}},\ \bibinfo {pages}
  {015005} (\bibinfo {year} {2018})}\BibitemShut {NoStop}%
\bibitem [{\citenamefont {{\v Z}elezn{\'y}}\ \emph {et~al.}(2018)\citenamefont
  {{\v Z}elezn{\'y}}, \citenamefont {Wadley}, \citenamefont {Olejn{\'i}k},
  \citenamefont {Hoffmann},\ and\ \citenamefont {Ohno}}]{zelezny2018}%
  \BibitemOpen
  \bibfield  {author} {\bibinfo {author} {\bibfnamefont {J.}~\bibnamefont {{\v
  Z}elezn{\'y}}}, \bibinfo {author} {\bibfnamefont {P.}~\bibnamefont {Wadley}},
  \bibinfo {author} {\bibfnamefont {K.}~\bibnamefont {Olejn{\'i}k}}, \bibinfo
  {author} {\bibfnamefont {A.}~\bibnamefont {Hoffmann}}, \ and\ \bibinfo
  {author} {\bibfnamefont {H.}~\bibnamefont {Ohno}},\ }\href {\doibase
  10.1038/s41567-018-0062-7} {\bibfield  {journal} {\bibinfo  {journal} {Nat.
  Phys.}\ }\textbf {\bibinfo {volume} {14}},\ \bibinfo {pages} {220} (\bibinfo
  {year} {2018})}\BibitemShut {NoStop}%
\bibitem [{\citenamefont {N{\'u}{\~n}ez}\ \emph {et~al.}(2006)\citenamefont
  {N{\'u}{\~n}ez}, \citenamefont {Duine}, \citenamefont {Haney},\ and\
  \citenamefont {MacDonald}}]{nunez2006}%
  \BibitemOpen
  \bibfield  {author} {\bibinfo {author} {\bibfnamefont {A.~S.}\ \bibnamefont
  {N{\'u}{\~n}ez}}, \bibinfo {author} {\bibfnamefont {R.~A.}\ \bibnamefont
  {Duine}}, \bibinfo {author} {\bibfnamefont {P.}~\bibnamefont {Haney}}, \ and\
  \bibinfo {author} {\bibfnamefont {A.~H.}\ \bibnamefont {MacDonald}},\ }\href
  {\doibase 10.1103/PhysRevB.73.214426} {\bibfield  {journal} {\bibinfo
  {journal} {Phys. Rev. B}\ }\textbf {\bibinfo {volume} {73}},\ \bibinfo
  {pages} {214426} (\bibinfo {year} {2006})}\BibitemShut {NoStop}%
\bibitem [{\citenamefont {Wei}\ \emph {et~al.}(2007)\citenamefont {Wei},
  \citenamefont {Sharma}, \citenamefont {Nunez}, \citenamefont {Haney},
  \citenamefont {Duine}, \citenamefont {Bass}, \citenamefont {MacDonald},\ and\
  \citenamefont {Tsoi}}]{wei2007}%
  \BibitemOpen
  \bibfield  {author} {\bibinfo {author} {\bibfnamefont {Z.}~\bibnamefont
  {Wei}}, \bibinfo {author} {\bibfnamefont {A.}~\bibnamefont {Sharma}},
  \bibinfo {author} {\bibfnamefont {A.~S.}\ \bibnamefont {Nunez}}, \bibinfo
  {author} {\bibfnamefont {P.~M.}\ \bibnamefont {Haney}}, \bibinfo {author}
  {\bibfnamefont {R.~A.}\ \bibnamefont {Duine}}, \bibinfo {author}
  {\bibfnamefont {J.}~\bibnamefont {Bass}}, \bibinfo {author} {\bibfnamefont
  {A.~H.}\ \bibnamefont {MacDonald}}, \ and\ \bibinfo {author} {\bibfnamefont
  {M.}~\bibnamefont {Tsoi}},\ }\href {\doibase 10.1103/PhysRevLett.98.116603}
  {\bibfield  {journal} {\bibinfo  {journal} {Phys. Rev. Lett.}\ }\textbf
  {\bibinfo {volume} {98}},\ \bibinfo {pages} {116603} (\bibinfo {year}
  {2007})}\BibitemShut {NoStop}%
\bibitem [{\citenamefont {Urazhdin}\ and\ \citenamefont
  {Anthony}(2007)}]{urazhdin2007}%
  \BibitemOpen
  \bibfield  {author} {\bibinfo {author} {\bibfnamefont {S.}~\bibnamefont
  {Urazhdin}}\ and\ \bibinfo {author} {\bibfnamefont {N.}~\bibnamefont
  {Anthony}},\ }\href {\doibase 10.1103/PhysRevLett.99.046602} {\bibfield
  {journal} {\bibinfo  {journal} {Phys. Rev. Lett.}\ }\textbf {\bibinfo
  {volume} {99}},\ \bibinfo {pages} {046602} (\bibinfo {year}
  {2007})}\BibitemShut {NoStop}%
\bibitem [{\citenamefont {Haney}\ and\ \citenamefont
  {MacDonald}(2008)}]{haney2008}%
  \BibitemOpen
  \bibfield  {author} {\bibinfo {author} {\bibfnamefont {P.~M.}\ \bibnamefont
  {Haney}}\ and\ \bibinfo {author} {\bibfnamefont {A.~H.}\ \bibnamefont
  {MacDonald}},\ }\href {\doibase 10.1103/PhysRevLett.100.196801} {\bibfield
  {journal} {\bibinfo  {journal} {Phys. Rev. Lett.}\ }\textbf {\bibinfo
  {volume} {100}},\ \bibinfo {pages} {196801} (\bibinfo {year}
  {2008})}\BibitemShut {NoStop}%
\bibitem [{\citenamefont {Gomonay}\ and\ \citenamefont
  {Loktev}(2010)}]{gomonay2010}%
  \BibitemOpen
  \bibfield  {author} {\bibinfo {author} {\bibfnamefont {H.~V.}\ \bibnamefont
  {Gomonay}}\ and\ \bibinfo {author} {\bibfnamefont {V.~M.}\ \bibnamefont
  {Loktev}},\ }\href {\doibase 10.1103/PhysRevB.81.144427} {\bibfield
  {journal} {\bibinfo  {journal} {Phys. Rev. B}\ }\textbf {\bibinfo {volume}
  {81}},\ \bibinfo {pages} {144427} (\bibinfo {year} {2010})}\BibitemShut
  {NoStop}%
\bibitem [{\citenamefont {Saidaoui}\ \emph {et~al.}(2014)\citenamefont
  {Saidaoui}, \citenamefont {Manchon},\ and\ \citenamefont
  {Waintal}}]{saidaoui2014}%
  \BibitemOpen
  \bibfield  {author} {\bibinfo {author} {\bibfnamefont {H.~B.~M.}\
  \bibnamefont {Saidaoui}}, \bibinfo {author} {\bibfnamefont {A.}~\bibnamefont
  {Manchon}}, \ and\ \bibinfo {author} {\bibfnamefont {X.}~\bibnamefont
  {Waintal}},\ }\href {\doibase 10.1103/PhysRevB.89.174430} {\bibfield
  {journal} {\bibinfo  {journal} {Phys. Rev. B}\ }\textbf {\bibinfo {volume}
  {89}},\ \bibinfo {pages} {174430} (\bibinfo {year} {2014})}\BibitemShut
  {NoStop}%
\bibitem [{\citenamefont {Cheng}\ \emph {et~al.}(2014)\citenamefont {Cheng},
  \citenamefont {Xiao}, \citenamefont {Niu},\ and\ \citenamefont
  {Brataas}}]{cheng2014}%
  \BibitemOpen
  \bibfield  {author} {\bibinfo {author} {\bibfnamefont {R.}~\bibnamefont
  {Cheng}}, \bibinfo {author} {\bibfnamefont {J.}~\bibnamefont {Xiao}},
  \bibinfo {author} {\bibfnamefont {Q.}~\bibnamefont {Niu}}, \ and\ \bibinfo
  {author} {\bibfnamefont {A.}~\bibnamefont {Brataas}},\ }\href {\doibase
  10.1103/PhysRevLett.113.057601} {\bibfield  {journal} {\bibinfo  {journal}
  {Phys. Rev. Lett.}\ }\textbf {\bibinfo {volume} {113}},\ \bibinfo {pages}
  {057601} (\bibinfo {year} {2014})}\BibitemShut {NoStop}%
\bibitem [{\citenamefont {Xu}\ \emph {et~al.}(2008)\citenamefont {Xu},
  \citenamefont {Wang},\ and\ \citenamefont {Xia}}]{xu2008}%
  \BibitemOpen
  \bibfield  {author} {\bibinfo {author} {\bibfnamefont {Y.}~\bibnamefont
  {Xu}}, \bibinfo {author} {\bibfnamefont {S.}~\bibnamefont {Wang}}, \ and\
  \bibinfo {author} {\bibfnamefont {K.}~\bibnamefont {Xia}},\ }\href {\doibase
  10.1103/PhysRevLett.100.226602} {\bibfield  {journal} {\bibinfo  {journal}
  {Phys. Rev. Lett.}\ }\textbf {\bibinfo {volume} {100}},\ \bibinfo {pages}
  {226602} (\bibinfo {year} {2008})}\BibitemShut {NoStop}%
\bibitem [{\citenamefont {Swaving}\ and\ \citenamefont
  {Duine}(2011)}]{swaving2011}%
  \BibitemOpen
  \bibfield  {author} {\bibinfo {author} {\bibfnamefont {A.~C.}\ \bibnamefont
  {Swaving}}\ and\ \bibinfo {author} {\bibfnamefont {R.~A.}\ \bibnamefont
  {Duine}},\ }\href {\doibase 10.1103/PhysRevB.83.054428} {\bibfield  {journal}
  {\bibinfo  {journal} {Phys. Rev. B}\ }\textbf {\bibinfo {volume} {83}},\
  \bibinfo {pages} {054428} (\bibinfo {year} {2011})}\BibitemShut {NoStop}%
\bibitem [{\citenamefont {Hals}\ \emph {et~al.}(2011)\citenamefont {Hals},
  \citenamefont {Tserkovnyak},\ and\ \citenamefont {Brataas}}]{hals2011}%
  \BibitemOpen
  \bibfield  {author} {\bibinfo {author} {\bibfnamefont {K.~M.~D.}\
  \bibnamefont {Hals}}, \bibinfo {author} {\bibfnamefont {Y.}~\bibnamefont
  {Tserkovnyak}}, \ and\ \bibinfo {author} {\bibfnamefont {A.}~\bibnamefont
  {Brataas}},\ }\href {\doibase 10.1103/PhysRevLett.106.107206} {\bibfield
  {journal} {\bibinfo  {journal} {Phys. Rev. Lett.}\ }\textbf {\bibinfo
  {volume} {106}},\ \bibinfo {pages} {107206} (\bibinfo {year}
  {2011})}\BibitemShut {NoStop}%
\bibitem [{\citenamefont {Tveten}\ \emph {et~al.}(2013)\citenamefont {Tveten},
  \citenamefont {Qaiumzadeh}, \citenamefont {Tretiakov},\ and\ \citenamefont
  {Brataas}}]{tveten2013}%
  \BibitemOpen
  \bibfield  {author} {\bibinfo {author} {\bibfnamefont {E.~G.}\ \bibnamefont
  {Tveten}}, \bibinfo {author} {\bibfnamefont {A.}~\bibnamefont {Qaiumzadeh}},
  \bibinfo {author} {\bibfnamefont {O.~A.}\ \bibnamefont {Tretiakov}}, \ and\
  \bibinfo {author} {\bibfnamefont {A.}~\bibnamefont {Brataas}},\ }\href
  {\doibase 10.1103/PhysRevLett.110.127208} {\bibfield  {journal} {\bibinfo
  {journal} {Phys. Rev. Lett.}\ }\textbf {\bibinfo {volume} {110}},\ \bibinfo
  {pages} {127208} (\bibinfo {year} {2013})}\BibitemShut {NoStop}%
\bibitem [{\citenamefont {Yamane}\ \emph {et~al.}(2016)\citenamefont {Yamane},
  \citenamefont {Ieda},\ and\ \citenamefont {Sinova}}]{yamane2016}%
  \BibitemOpen
  \bibfield  {author} {\bibinfo {author} {\bibfnamefont {Y.}~\bibnamefont
  {Yamane}}, \bibinfo {author} {\bibfnamefont {J.}~\bibnamefont {Ieda}}, \ and\
  \bibinfo {author} {\bibfnamefont {J.}~\bibnamefont {Sinova}},\ }\href
  {\doibase 10.1103/PhysRevB.94.054409} {\bibfield  {journal} {\bibinfo
  {journal} {Phys. Rev. B}\ }\textbf {\bibinfo {volume} {94}},\ \bibinfo
  {pages} {054409} (\bibinfo {year} {2016})}\BibitemShut {NoStop}%
\bibitem [{\citenamefont {Barker}\ and\ \citenamefont
  {Tretiakov}(2016)}]{barker2016}%
  \BibitemOpen
  \bibfield  {author} {\bibinfo {author} {\bibfnamefont {J.}~\bibnamefont
  {Barker}}\ and\ \bibinfo {author} {\bibfnamefont {O.~A.}\ \bibnamefont
  {Tretiakov}},\ }\href {\doibase 10.1103/PhysRevLett.116.147203} {\bibfield
  {journal} {\bibinfo  {journal} {Phys. Rev. Lett.}\ }\textbf {\bibinfo
  {volume} {116}},\ \bibinfo {pages} {147203} (\bibinfo {year}
  {2016})}\BibitemShut {NoStop}%
\bibitem [{\citenamefont {Park}\ \emph {et~al.}(2020)\citenamefont {Park},
  \citenamefont {Jeong}, \citenamefont {Oh}, \citenamefont {Go}, \citenamefont
  {Oh}, \citenamefont {Kim}, \citenamefont {Lee},\ and\ \citenamefont
  {Lee}}]{park2020}%
  \BibitemOpen
  \bibfield  {author} {\bibinfo {author} {\bibfnamefont {H.-J.}\ \bibnamefont
  {Park}}, \bibinfo {author} {\bibfnamefont {Y.}~\bibnamefont {Jeong}},
  \bibinfo {author} {\bibfnamefont {S.-H.}\ \bibnamefont {Oh}}, \bibinfo
  {author} {\bibfnamefont {G.}~\bibnamefont {Go}}, \bibinfo {author}
  {\bibfnamefont {J.~H.}\ \bibnamefont {Oh}}, \bibinfo {author} {\bibfnamefont
  {K.-W.}\ \bibnamefont {Kim}}, \bibinfo {author} {\bibfnamefont {H.-W.}\
  \bibnamefont {Lee}}, \ and\ \bibinfo {author} {\bibfnamefont {K.-J.}\
  \bibnamefont {Lee}},\ }\href {\doibase 10.1103/PhysRevB.101.144431}
  {\bibfield  {journal} {\bibinfo  {journal} {Phys. Rev. B}\ }\textbf {\bibinfo
  {volume} {101}},\ \bibinfo {pages} {144431} (\bibinfo {year}
  {2020})}\BibitemShut {NoStop}%
\bibitem [{\citenamefont {Nakane}\ \emph {et~al.}(2020)\citenamefont {Nakane},
  \citenamefont {Nakazawa},\ and\ \citenamefont {Kohno}}]{nakane2020}%
  \BibitemOpen
  \bibfield  {author} {\bibinfo {author} {\bibfnamefont {J.~J.}\ \bibnamefont
  {Nakane}}, \bibinfo {author} {\bibfnamefont {K.}~\bibnamefont {Nakazawa}}, \
  and\ \bibinfo {author} {\bibfnamefont {H.}~\bibnamefont {Kohno}},\ }\href
  {\doibase 10.1103/PhysRevB.101.174432} {\bibfield  {journal} {\bibinfo
  {journal} {Phys. Rev. B}\ }\textbf {\bibinfo {volume} {101}},\ \bibinfo
  {pages} {174432} (\bibinfo {year} {2020})}\BibitemShut {NoStop}%
\bibitem [{\citenamefont {Fuji}\ \emph {et~al.}(2015)\citenamefont {Fuji},
  \citenamefont {Pollmann},\ and\ \citenamefont {Oshikawa}}]{fuji2015}%
  \BibitemOpen
  \bibfield  {author} {\bibinfo {author} {\bibfnamefont {Y.}~\bibnamefont
  {Fuji}}, \bibinfo {author} {\bibfnamefont {F.}~\bibnamefont {Pollmann}}, \
  and\ \bibinfo {author} {\bibfnamefont {M.}~\bibnamefont {Oshikawa}},\ }\href
  {\doibase 10.1103/PhysRevLett.114.177204} {\bibfield  {journal} {\bibinfo
  {journal} {Phys. Rev. Lett.}\ }\textbf {\bibinfo {volume} {114}},\ \bibinfo
  {pages} {177204} (\bibinfo {year} {2015})}\BibitemShut {NoStop}%
\bibitem [{\citenamefont {Cheng}\ and\ \citenamefont {Niu}(2012)}]{cheng2012}%
  \BibitemOpen
  \bibfield  {author} {\bibinfo {author} {\bibfnamefont {R.}~\bibnamefont
  {Cheng}}\ and\ \bibinfo {author} {\bibfnamefont {Q.}~\bibnamefont {Niu}},\
  }\href {\doibase 10.1103/PhysRevB.86.245118} {\bibfield  {journal} {\bibinfo
  {journal} {Phys. Rev. B}\ }\textbf {\bibinfo {volume} {86}},\ \bibinfo
  {pages} {245118} (\bibinfo {year} {2012})}\BibitemShut {NoStop}%
\bibitem [{\citenamefont {Korenman}\ \emph {et~al.}(1977)\citenamefont
  {Korenman}, \citenamefont {Murray},\ and\ \citenamefont
  {Prange}}]{korenman1977}%
  \BibitemOpen
  \bibfield  {author} {\bibinfo {author} {\bibfnamefont {V.}~\bibnamefont
  {Korenman}}, \bibinfo {author} {\bibfnamefont {J.~L.}\ \bibnamefont
  {Murray}}, \ and\ \bibinfo {author} {\bibfnamefont {R.~E.}\ \bibnamefont
  {Prange}},\ }\href {\doibase 10.1103/PhysRevB.16.4032} {\bibfield  {journal}
  {\bibinfo  {journal} {Phys. Rev. B}\ }\textbf {\bibinfo {volume} {16}},\
  \bibinfo {pages} {4032} (\bibinfo {year} {1977})}\BibitemShut {NoStop}%
\bibitem [{\citenamefont {Bazaliy}\ \emph {et~al.}(1998)\citenamefont
  {Bazaliy}, \citenamefont {Jones},\ and\ \citenamefont {Zhang}}]{bazaliy1998}%
  \BibitemOpen
  \bibfield  {author} {\bibinfo {author} {\bibfnamefont {Y.~B.}\ \bibnamefont
  {Bazaliy}}, \bibinfo {author} {\bibfnamefont {B.~A.}\ \bibnamefont {Jones}},
  \ and\ \bibinfo {author} {\bibfnamefont {S.-C.}\ \bibnamefont {Zhang}},\
  }\href {\doibase 10.1103/PhysRevB.57.R3213} {\bibfield  {journal} {\bibinfo
  {journal} {Phys. Rev. B}\ }\textbf {\bibinfo {volume} {57}},\ \bibinfo
  {pages} {R3213} (\bibinfo {year} {1998})}\BibitemShut {NoStop}%
\bibitem [{\citenamefont {Kohno}\ and\ \citenamefont
  {Shibata}(2007)}]{kohno2007}%
  \BibitemOpen
  \bibfield  {author} {\bibinfo {author} {\bibfnamefont {H.}~\bibnamefont
  {Kohno}}\ and\ \bibinfo {author} {\bibfnamefont {J.}~\bibnamefont
  {Shibata}},\ }\href {\doibase 10.1143/JPSJ.76.063710} {\bibfield  {journal}
  {\bibinfo  {journal} {J. Phys. Soc. Jpn.}\ }\textbf {\bibinfo {volume}
  {76}},\ \bibinfo {pages} {063710} (\bibinfo {year} {2007})}\BibitemShut
  {NoStop}%
\bibitem [{STT()}]{STTSM}%
  \BibitemOpen
  \href@noop {} {\bibinfo  {journal} {See Supplemental Material at [URL will be
  inserted by publisher] for details of calculations}\ }\BibitemShut {NoStop}%
\bibitem [{\citenamefont {Hayami}\ \emph {et~al.}(2020)\citenamefont {Hayami},
  \citenamefont {Yanagi},\ and\ \citenamefont {Kusunose}}]{hayami2020}%
  \BibitemOpen
\bibfield  {journal} {  }\bibfield  {author} {\bibinfo {author} {\bibfnamefont
  {S.}~\bibnamefont {Hayami}}, \bibinfo {author} {\bibfnamefont
  {Y.}~\bibnamefont {Yanagi}}, \ and\ \bibinfo {author} {\bibfnamefont
  {H.}~\bibnamefont {Kusunose}},\ }\href {\doibase 10.1103/PhysRevB.101.220403}
  {\bibfield  {journal} {\bibinfo  {journal} {Phys. Rev. B}\ }\textbf {\bibinfo
  {volume} {101}},\ \bibinfo {pages} {220403} (\bibinfo {year}
  {2020})}\BibitemShut {NoStop}%
\bibitem [{\citenamefont {Yuan}\ \emph {et~al.}(2020)\citenamefont {Yuan},
  \citenamefont {Wang}, \citenamefont {Luo}, \citenamefont {Rashba},\ and\
  \citenamefont {Zunger}}]{yuan2020}%
  \BibitemOpen
  \bibfield  {author} {\bibinfo {author} {\bibfnamefont {L.-D.}\ \bibnamefont
  {Yuan}}, \bibinfo {author} {\bibfnamefont {Z.}~\bibnamefont {Wang}}, \bibinfo
  {author} {\bibfnamefont {J.-W.}\ \bibnamefont {Luo}}, \bibinfo {author}
  {\bibfnamefont {E.~I.}\ \bibnamefont {Rashba}}, \ and\ \bibinfo {author}
  {\bibfnamefont {A.}~\bibnamefont {Zunger}},\ }\href {\doibase
  10.1103/PhysRevB.102.014422} {\bibfield  {journal} {\bibinfo  {journal}
  {Phys. Rev. B}\ }\textbf {\bibinfo {volume} {102}},\ \bibinfo {pages}
  {014422} (\bibinfo {year} {2020})}\BibitemShut {NoStop}%
\bibitem [{\citenamefont {Onoda}\ \emph {et~al.}(2004)\citenamefont {Onoda},
  \citenamefont {Tatara},\ and\ \citenamefont {Nagaosa}}]{onoda2004}%
  \BibitemOpen
  \bibfield  {author} {\bibinfo {author} {\bibfnamefont {M.}~\bibnamefont
  {Onoda}}, \bibinfo {author} {\bibfnamefont {G.}~\bibnamefont {Tatara}}, \
  and\ \bibinfo {author} {\bibfnamefont {N.}~\bibnamefont {Nagaosa}},\ }\href
  {\doibase 10.1143/JPSJ.73.2624} {\bibfield  {journal} {\bibinfo  {journal}
  {J. Phys. Soc. Jpn.}\ }\textbf {\bibinfo {volume} {73}},\ \bibinfo {pages}
  {2624} (\bibinfo {year} {2004})}\BibitemShut {NoStop}%
\bibitem [{\citenamefont {Sachdev}(2011)}]{sachdev2011}%
  \BibitemOpen
  \bibfield  {author} {\bibinfo {author} {\bibfnamefont {S.}~\bibnamefont
  {Sachdev}},\ }\href {\doibase 10.1017/CBO9780511973765} {\emph {\bibinfo
  {title} {Quantum {{Phase Transitions}}}}},\ \bibinfo {edition} {2nd}\ ed.\
  (\bibinfo  {publisher} {{Cambridge University Press}},\ \bibinfo {address}
  {{Cambridge}},\ \bibinfo {year} {2011})\BibitemShut {NoStop}%
\bibitem [{\citenamefont {Tatara}\ \emph {et~al.}(2008)\citenamefont {Tatara},
  \citenamefont {Kohno},\ and\ \citenamefont {Shibata}}]{tatara2008}%
  \BibitemOpen
  \bibfield  {author} {\bibinfo {author} {\bibfnamefont {G.}~\bibnamefont
  {Tatara}}, \bibinfo {author} {\bibfnamefont {H.}~\bibnamefont {Kohno}}, \
  and\ \bibinfo {author} {\bibfnamefont {J.}~\bibnamefont {Shibata}},\ }\href
  {\doibase 10.1016/j.physrep.2008.07.003} {\bibfield  {journal} {\bibinfo
  {journal} {Phys. Rep.}\ }\textbf {\bibinfo {volume} {468}},\ \bibinfo {pages}
  {213} (\bibinfo {year} {2008})}\BibitemShut {NoStop}%
\bibitem [{\citenamefont {Fujimoto}\ and\ \citenamefont
  {Matsuo}(2019)}]{fujimoto2019a}%
  \BibitemOpen
  \bibfield  {author} {\bibinfo {author} {\bibfnamefont {J.}~\bibnamefont
  {Fujimoto}}\ and\ \bibinfo {author} {\bibfnamefont {M.}~\bibnamefont
  {Matsuo}},\ }\href {\doibase 10.1103/PhysRevB.100.220402} {\bibfield
  {journal} {\bibinfo  {journal} {Phys. Rev. B}\ }\textbf {\bibinfo {volume}
  {100}},\ \bibinfo {pages} {220402} (\bibinfo {year} {2019})}\BibitemShut
  {NoStop}%
\bibitem [{\citenamefont {Papanicolaou}(1995)}]{papanicolaou1995}%
  \BibitemOpen
  \bibfield  {author} {\bibinfo {author} {\bibfnamefont {N.}~\bibnamefont
  {Papanicolaou}},\ }\href {\doibase 10.1103/PhysRevB.51.15062} {\bibfield
  {journal} {\bibinfo  {journal} {Phys. Rev. B}\ }\textbf {\bibinfo {volume}
  {51}},\ \bibinfo {pages} {15062} (\bibinfo {year} {1995})}\BibitemShut
  {NoStop}%
\bibitem [{\citenamefont {Jaramillo}\ \emph {et~al.}(2007)\citenamefont
  {Jaramillo}, \citenamefont {Rosenbaum}, \citenamefont {Isaacs}, \citenamefont
  {Shpyrko}, \citenamefont {Evans}, \citenamefont {Aeppli},\ and\ \citenamefont
  {Cai}}]{jaramillo2007}%
  \BibitemOpen
  \bibfield  {author} {\bibinfo {author} {\bibfnamefont {R.}~\bibnamefont
  {Jaramillo}}, \bibinfo {author} {\bibfnamefont {T.~F.}\ \bibnamefont
  {Rosenbaum}}, \bibinfo {author} {\bibfnamefont {E.~D.}\ \bibnamefont
  {Isaacs}}, \bibinfo {author} {\bibfnamefont {O.~G.}\ \bibnamefont {Shpyrko}},
  \bibinfo {author} {\bibfnamefont {P.~G.}\ \bibnamefont {Evans}}, \bibinfo
  {author} {\bibfnamefont {G.}~\bibnamefont {Aeppli}}, \ and\ \bibinfo {author}
  {\bibfnamefont {Z.}~\bibnamefont {Cai}},\ }\href {\doibase
  10.1103/PhysRevLett.98.117206} {\bibfield  {journal} {\bibinfo  {journal}
  {Phys. Rev. Lett.}\ }\textbf {\bibinfo {volume} {98}},\ \bibinfo {pages}
  {117206} (\bibinfo {year} {2007})}\BibitemShut {NoStop}%
\bibitem [{\citenamefont {Okuno}\ \emph {et~al.}(2019)\citenamefont {Okuno},
  \citenamefont {Kim}, \citenamefont {Oh}, \citenamefont {Kim}, \citenamefont
  {Hirata}, \citenamefont {Nishimura}, \citenamefont {Ham}, \citenamefont
  {Futakawa}, \citenamefont {Yoshikawa}, \citenamefont {Tsukamoto},
  \citenamefont {Tserkovnyak}, \citenamefont {Shiota}, \citenamefont
  {Moriyama}, \citenamefont {Kim}, \citenamefont {Lee},\ and\ \citenamefont
  {Ono}}]{okuno2019}%
  \BibitemOpen
  \bibfield  {author} {\bibinfo {author} {\bibfnamefont {T.}~\bibnamefont
  {Okuno}}, \bibinfo {author} {\bibfnamefont {D.-H.}\ \bibnamefont {Kim}},
  \bibinfo {author} {\bibfnamefont {S.-H.}\ \bibnamefont {Oh}}, \bibinfo
  {author} {\bibfnamefont {S.~K.}\ \bibnamefont {Kim}}, \bibinfo {author}
  {\bibfnamefont {Y.}~\bibnamefont {Hirata}}, \bibinfo {author} {\bibfnamefont
  {T.}~\bibnamefont {Nishimura}}, \bibinfo {author} {\bibfnamefont {W.~S.}\
  \bibnamefont {Ham}}, \bibinfo {author} {\bibfnamefont {Y.}~\bibnamefont
  {Futakawa}}, \bibinfo {author} {\bibfnamefont {H.}~\bibnamefont {Yoshikawa}},
  \bibinfo {author} {\bibfnamefont {A.}~\bibnamefont {Tsukamoto}}, \bibinfo
  {author} {\bibfnamefont {Y.}~\bibnamefont {Tserkovnyak}}, \bibinfo {author}
  {\bibfnamefont {Y.}~\bibnamefont {Shiota}}, \bibinfo {author} {\bibfnamefont
  {T.}~\bibnamefont {Moriyama}}, \bibinfo {author} {\bibfnamefont {K.-J.}\
  \bibnamefont {Kim}}, \bibinfo {author} {\bibfnamefont {K.-J.}\ \bibnamefont
  {Lee}}, \ and\ \bibinfo {author} {\bibfnamefont {T.}~\bibnamefont {Ono}},\
  }\href {\doibase 10.1038/s41928-019-0303-5} {\bibfield  {journal} {\bibinfo
  {journal} {Nat. Electron.}\ }\textbf {\bibinfo {volume} {2}},\ \bibinfo
  {pages} {389} (\bibinfo {year} {2019})}\BibitemShut {NoStop}%
\end{thebibliography}%

\onecolumngrid
\end{document}


%
%
%
%
%
\title{Supplemental Material: \\Adiabatic and Nonadiabatic Spin-transfer Torques in Antiferromagnets}
\date{\today}
\author{Junji Fujimoto}
\affiliation{Kavli Institute for Theoretical Sciences, University of Chinese Academy of Sciences, Beijing, 100190, China}

\maketitle
\onecolumngrid
\tableofcontents

\section{\label{sec:Vk}Diagonalization matrix}
In this section, we show the explicit form of the diagonalization matrix $V_{\bk}$.
First, we introduce the following vector form
\begin{align}
\bm{X}_{\bk}
	& = \begin{pmatrix}
		\Re \eta_{\bk}  \sigma^{0}
	\\	\Im \eta_{\bk} \sigma^{0}
	\\	\mathcal{J}_{sd} \sigma^{z}
	\end{pmatrix}
	= \Delta_{\bk} \begin{pmatrix}
		\sigma^{0} \sin \vartheta_{\bk} \cos \varphi_{\bk}
	\\	\sigma^{0} \sin \vartheta_{\bk} \sin \varphi_{\bk}
	\\	\sigma^z \cos \vartheta_{\bk}
	\end{pmatrix}
,\end{align}
and hence the Hamiltonian is expressed as $H_{\bk} = T_{\bk} \rho_0 \sigma^0 - \bm{X}_{\bk} \cdot \bm{\rho}$.
Here, we define the $V_{\bk}$ as
\begin{align}
V_{\bk}^{}
	& = \tilde{\bm{\chi}}_{\bk} \cdot \bm{\rho}
,\end{align}
where
\begin{align}
\tilde{\bm{\chi}}_{\bk}
	& = \begin{pmatrix}
		\sigma^0 \sin (\vartheta_{\bk}/2) \cos \varphi_{\bk}
	\\	\sigma^0 \sin (\vartheta_{\bk}/2) \sin \varphi_{\bk}
	\\	\sigma^z \cos (\vartheta_{\bk}/2)
	\end{pmatrix}
	= \bm{\chi}_{\bk}^{\perp} \sigma^{0} + \chi_{\bk}^{z} \hat{z} \sigma^{z}
\label{eq:chi}
.\end{align}
By using $V_{\bk}$, one obtains
\begin{align}
V_{\bk}^{\dagger} \bigl( \bm{X}_{\bk} \cdot \bm{\rho} \bigr) V_{\bk}^{}
	& = \Delta_{\bk} (\bm{\chi}_{\bk}^{\perp} \cdot \bm{\rho}^{\perp} \sigma^0 + \chi_{\bk}^{z} \rho_3 \sigma^z)
		\bigl( \sin \vartheta_{\bk} (\hat{e}_{\varphi_{\bk}} \cdot \bm{\rho}^{\perp}) \sigma^0 + \cos \vartheta_{\bk} \rho_3 \sigma^z \bigr)
		(\bm{\chi}_{\bk}^{\perp} \cdot \bm{\rho}^{\perp} \sigma^0 + \chi_{\bk}^{z} \rho_3 \sigma^z)
\notag \\
	& = \Delta_{\bk} \Bigl( \cos (\vartheta_{\bk}/2) \rho_0 \sigma^0 + \zi \sin (\vartheta_{\bk}/2) (\hat{z} \times \hat{e}_{\varphi_{\bk}}) \cdot \bm{\rho}^{\perp} \sigma^z
		\Bigr)
		(\bm{\chi}_{\bk}^{\perp} \cdot \bm{\rho}^{\perp} \sigma^0 + \chi_{\bk}^{z} \rho_3 \sigma^z)
\notag \\
	& = \Delta_{\bk} \rho_3 \sigma^z
,\end{align}
where $\hat{e}_{\varphi_{\bk}} = (\cos \varphi_{\bk}, \sin \varphi_{\bk}, 0)$.
Hence, we have $V^{\dagger}_{\bk} H_{\bk} V_{\bk} = T_{\bk} \rho_0 \sigma^0 - \Delta_{\bk} \rho_3 \sigma^z$.

\section{\label{sec:matrix}Projection operator and projected spaces}
Here, we show the projection operator into the eigenstates and the projected spaces.
The projection operator into the bonding state ($s = +$) and into the antibonding state ($s = -$) are given by
\begin{align}
P_{s}
	& = \frac{\rho_0 \sigma^0 + s \rho_3 \sigma^z}{2}
,\end{align}
or, in the matrix form,
\begin{align}
P_{+}
	= \begin{pmatrix}
		1
	&	0
	&	0
	&	0
	\\	0
	&	0
	&	0
	&	0
	\\	0
	&	0
	&	0
	&	0
	\\	0
	&	0
	&	0
	&	1
	\end{pmatrix}
, & \qquad
P_{-}
	= \begin{pmatrix}
		0
	&	0
	&	0
	&	0
	\\	0
	&	1
	&	0
	&	0
	\\	0
	&	0
	&	1
	&	0
	\\	0
	&	0
	&	0
	&	0
	\end{pmatrix}
.\end{align}
Since the operator $P_s$ ($s = \pm$) holds the following relation,
\begin{align}
P_{s}^2
	= P_{s}
\label{eq:idempotence_bond}
,\end{align}
the operator $P_{s}$ is a projection operator.
We also see
\begin{align}
P_{+} + P_{-}
	& = \rho_0 \sigma^0
\label{eq:orthogonality_bond}
,\end{align}
so that we have the orthogonality:
\begin{align}
P_{+} P_{-}
	= P_{-} P_{+}
	= 0
.\end{align}

Next, we derive the formulas for the projections of $\rho_{\mu} \sigma^{\nu}$ ($\mu = 0, 1, 2, 3$ and $\nu = 0, x, y, z$). 
We first divide $P_{s} \rho_{\mu} \sigma^{\nu}$ int two classes:
one is commutative, $P_{s} \rho_{\mu} \sigma^{\nu} = \rho_{\mu} \sigma^{\nu} P_{s}$ for $(\mu, \nu) = (0,0), (0,z), (3,0), (3,z), (1,x), (1,y), (2,x), (2,y)$, and the other is noncommutative $P_{s} \rho_{\mu} \sigma^{\nu} = \rho_{\mu} \sigma^{\nu} P_{\bar{s}}$ for $(\mu, \nu) = (0,x), (0,y), (1,0), (1,z), (2,0), (2,z), (3,x), (3,y)$, where $\bar{s} = - s$.
The commutative class is equivalent to the adiabatic contribution;
\begin{align}
\sum_{\bk} (V_{\bk} \Psi_{\bk})^{\dagger} \rho_{\mu} \sigma^{\nu} V_{\bk} \Psi_{\bk}
	& = \sum_{\bk} \sum_{s,s'} (P_s V_{\bk} \Psi_{\bk})^{\dagger} \left( P_s \rho_{\mu} \sigma^{\nu} P_{s'} \right) P_{s'} V_{\bk} \Psi_{\bk}
\notag \\
	& = \sum_{\bk} \sum_{s,s'} \psi_{\bk, s}^{\dagger} \left( P_s \rho_{\mu} \sigma^{\nu} P_{s'} \right) \psi_{\bk, s'}
\notag \\
	& = \sum_{\bk} \sum_{s,s'} \psi_{\bk, s}^{\dagger} \left( P_{s} P_{s'} \rho_{\mu} \sigma^{\nu} \right) \psi_{\bk, s'}
\notag \\
	& = \sum_{\bk} \sum_{s} \psi_{\bk, s}^{\dagger} \left( P_{s} \rho_{\mu} \sigma^{\nu} \right) \psi_{\bk, s}
,\end{align}
where we inserted $1 = \rho_0 \sigma_0 = \sum_{s = \pm} P_{s}$ and use $P_s = (P_s)^2$ in the first equal, and $P_s V_{\bk} \Psi_{\bk} = \psi_{\bk, s}$ is the eigenstate operator.
The similar calculation shows that the noncommutative class is equivalent to the nonadiabatic contribution, which is a mixing of the bonding and antibonding states,
\begin{align}
\sum_{\bk} (V_{\bk} \Psi_{\bk})^{\dagger} \rho_{\mu} \sigma^{\nu} V_{\bk} \Psi_{\bk}
	& = \sum_{\bk} \sum_{s,s'} \psi_{\bk, s}^{\dagger} \left( P_s \rho_{\mu} \sigma^{\nu} P_{s'} \right) \psi_{\bk, s'}
\notag \\
	& = \sum_{\bk} \sum_{s,s'} \psi_{\bk, s}^{\dagger} \left( P_{s} P_{\bar{s}'} \rho_{\mu} \sigma^{\nu} \right) \psi_{\bk, s'}
\notag \\
	& = \sum_{\bk} \sum_{s} \psi_{\bk, s}^{\dagger} \left( P_{s} \rho_{\mu} \sigma^{\nu} \right) \psi_{\bk, \bar{s}}
.\end{align}

Then, we show the explicit forms for each class.
\subsection{Commutative class}
For the commutative class $(\mu, \nu) = (0,0), (0,z), (3,0), (3,z), (1,x), (1,y), (2,x), (2,y)$, we have $P_{+} \rho_{\mu} \sigma^{\nu}$, 
\begin{subequations}
\begin{align}
P_{+} \rho_0 \sigma^{0}
	= P_{+} \rho_3 \sigma^{z}
	& = \begin{pmatrix}
		+1
	&
	&
	&	0
	\\	
	&
	&
	&
	\\	
	&
	&
	&
	\\	0
	&
	&
	&	+1
	\end{pmatrix}
	\equiv \tau^{0}_{++}
, \\
P_{+} \rho_1 \sigma^{x}
	= P_{+} (\zi \rho_2) (\zi \sigma^{y})
	& = \begin{pmatrix}
		0
	&
	&
	&	+1
	\\	
	&
	&
	&
	\\	
	&
	&
	&
	\\	+1
	&
	&
	&	0
	\end{pmatrix}
	\equiv \tau^{x}_{++}
, \\
P_{+} \rho_1 (\zi \sigma^{y})
	= P_{+} (\zi \rho_2) \sigma^{x}
	& = \zi \begin{pmatrix}
		0
	&
	&
	&	-\zi
	\\	
	&
	&
	&
	\\	
	&
	&
	&
	\\	+\zi
	&
	&
	&	0
	\end{pmatrix}
	\equiv \zi \tau^{y}_{++}
, \\
P_{+} \rho_3 \sigma^{0}
	= P_{+} \rho_0 \sigma^{z}
	& = \begin{pmatrix}
		+1
	&
	&
	&	0
	\\	
	&
	&
	&
	\\	
	&
	&
	&
	\\	0
	&
	&
	&	-1
	\end{pmatrix}
	\equiv \tau^{z}_{++}
,\end{align}
\end{subequations}
and $P_{-} \rho_{\mu} \sigma^{\nu}$,
\begin{subequations}
\begin{align}
P_{-} \rho_0 \sigma^{0}
	= \begin{pmatrix}
		\
	&
	&
	&
	\\	
	&	+1
	&	0
	&
	\\	
	&	0
	&	+1
	&
	\\
	&
	&
	&
	\end{pmatrix}
	= + \tau^0_{--}
, & \qquad
P_{-} \rho_3 \sigma^{z}
	= \begin{pmatrix}
		\
	&
	&
	&
	\\	
	&	-1
	&	0
	&
	\\	
	&	0
	&	-1
	&
	\\
	&
	&
	&
	\end{pmatrix}
	= - \tau^0_{--}
, \\ 
P_{-} \rho_1 \sigma^{x}
	= \begin{pmatrix}
		\
	&
	&
	&
	\\	
	&	0
	&	+1
	&
	\\	
	&	+1
	&	0
	&
	\\
	&
	&
	&
	\end{pmatrix}
	\equiv + \tau^{x}_{--}
, & \qquad
P_{-} (\zi \rho_2) (\zi \sigma^{y})
	= \begin{pmatrix}
		\
	&
	&
	&
	\\	
	&	0
	&	-1
	&
	\\	
	&	-1
	&	0
	&
	\\
	&
	&
	&
	\end{pmatrix}
	\equiv - \tau^{x}_{--}
, \\ 
P_{-} (\zi \rho_2) \sigma^{x}
	= \zi \begin{pmatrix}
		\
	&
	&
	&
	\\	
	&	0
	&	-\zi
	&
	\\	
	&	+\zi
	&	0
	&
	\\
	&
	&
	&
	\end{pmatrix}
	\equiv + \zi \tau^{y}_{--}
, & \qquad
P_{-} \rho_1 (\zi \sigma^{y})
	= \zi \begin{pmatrix}
		\
	&
	&
	&
	\\	
	&	0
	&	+\zi
	&
	\\	
	&	-\zi
	&	0
	&
	\\
	&
	&
	&
	\end{pmatrix}
	\equiv - \zi \tau^{y}_{--}
, \\ 
P_{-} \rho_3 \sigma^{0}
	= \begin{pmatrix}
		\
	&
	&
	&
	\\	
	&	+1
	&	0
	&
	\\	
	&	0
	&	-1
	&
	\\
	&
	&
	&
	\end{pmatrix}
	\equiv + \tau^{z}_{--}
, & \qquad
P_{-} \rho_0 \sigma^{z}
	= \begin{pmatrix}
		\
	&
	&
	&
	\\	
	&	-1
	&	0
	&
	\\	
	&	0
	&	+1
	&
	\\
	&
	&
	&
	\end{pmatrix}
	\equiv - \tau^{z}_{--}
.\end{align}
\end{subequations}
Here, we defined $\tau_{ss}^{\mu}$ ($s = \pm$ and $\mu = 0, x, y, z$).

To summarize these, we have
\begin{subequations}
\begin{align}
P_{s} \rho_{\mu=0,3} \sigma^0
	& = \tau^{\mu=0,z}_{ss}
, \\
P_{s} \bm{\rho}^{\perp} \sigma^x
	& = \bm{\tau}^{\perp}_{ss}
, \\
P_{s} \bm{\rho}^{\perp} \sigma^y
	& = - s \hat{z} \times \bm{\tau}^{\perp}_{ss}
\notag \\
	& = - \zi s \tau^z_{ss} \bm{\tau}^{\perp}_{ss}
, \\
P_{s} \rho_{\mu=0,3} \sigma^z
	& = s \tau^{\mu=z,0}_{ss}
.\end{align}
\label{eqs:P_eta-commutative}%
\end{subequations}

\subsection{Noncommutative class}
For the noncommutative class $(\mu, \nu) = (0,x), (0,y), (1,0), (1,z), (2,0), (2,z), (3,x), (3,y)$, we have $P_{+} \rho_{\mu} \sigma^{\nu}$,
\begin{subequations}
\begin{align}
P_{+} \rho_1 \sigma^0
	= P_{+} (\zi \rho_2) \sigma^z
	& = \begin{pmatrix}
		\
	&
	&	+1
	&
	\\
	&
	&
	&	0
	\\	0
	&
	&
	&
	\\
	&	+1
	&
	&
	\end{pmatrix}
	\equiv \tau_{+-}^{x}
, \\
P_{+} (\zi \rho_2) \sigma^0
	= P_{+} \rho_1 \sigma^z
	& = \begin{pmatrix}
		\
	&
	&	+1
	&
	\\
	&
	&
	&	0
	\\	0
	&
	&
	&
	\\
	&	-1
	&
	&
	\end{pmatrix}
	\equiv \zi \tau_{+-}^{y}
, \\
P_{+} \rho_0 \sigma^x
	= P_{+} \rho_3 (\zi \sigma^y)
	& = \begin{pmatrix}
		\
	&	+1
	&
	&
	\\	0
	&
	&
	&
	\\
	&
	&
	&	0
	\\
	&
	&	+1
	&
	\end{pmatrix}
	\equiv \tau_{+-}^{0}
, \\
P_{+} \rho_3 \sigma^x
	= P_{+} \rho_0 (\zi \sigma^y)
	& = \begin{pmatrix}
		\
	&	+1
	&
	&
	\\	0
	&
	&
	&
	\\
	&
	&
	&	0
	\\
	&
	&	-1
	&
	\end{pmatrix}
	\equiv \tau_{+-}^{z}
,\end{align}
\end{subequations}
and $P_{-} \rho_{\mu} \sigma^{\nu}$,
\begin{subequations}
\begin{align}
P_{-} \rho_1 \sigma^0
	= \begin{pmatrix}
		\
	&
	&	0
	&
	\\
	&
	&
	&	+1
	\\	+1
	&
	&
	&
	\\
	&	0
	&
	&
	\end{pmatrix}
	\equiv + \tau_{-+}^{x}
, & \qquad
P_{-} (\zi \rho_2) \sigma^z
	= \begin{pmatrix}
		\
	&
	&	0
	&
	\\
	&
	&
	&	-1
	\\	-1
	&
	&
	&
	\\
	&	0
	&
	&
	\end{pmatrix}
	\equiv - \tau_{-+}^{x}
, \\
P_{-} (\zi \rho_2) \sigma^0
	= \begin{pmatrix}
		\
	&
	&	0
	&
	\\
	&
	&
	&	+1
	\\	-1
	&
	&
	&
	\\
	&	0
	&
	&
	\end{pmatrix}
	\equiv + \zi \tau_{-+}^{y}
, & \qquad
P_{-} \rho_1 \sigma^z
	= \begin{pmatrix}
		\
	&
	&	0
	&
	\\
	&
	&
	&	-1
	\\	+1
	&
	&
	&
	\\
	&	0
	&
	&
	\end{pmatrix}
	\equiv - \zi \tau_{-+}^{y}
, \\
P_{-} \rho_0 \sigma^x
	= \begin{pmatrix}
		\
	&	0
	&
	&
	\\	+1
	&
	&
	&
	\\
	&
	&
	&	+1
	\\
	&
	&	0
	&
	\end{pmatrix}
	\equiv + \tau_{-+}^{0}
, & \qquad
P_{-} \rho_3 (\zi \sigma^y)
	= \begin{pmatrix}
		\
	&	0
	&
	&
	\\	-1
	&
	&
	&
	\\
	&
	&
	&	-1
	\\
	&
	&	0
	&
	\end{pmatrix}
	\equiv - \tau_{-+}^{0}
, \\
P_{-} \rho_3 \sigma^x
	= \begin{pmatrix}
		\
	&	0
	&
	&
	\\	+1
	&
	&
	&
	\\
	&
	&
	&	-1
	\\
	&
	&	0
	&
	\end{pmatrix}
	\equiv + \tau_{-+}^{z}
, & \qquad
P_{-} \rho_0 (\zi \sigma^y)
	= \begin{pmatrix}
		\
	&	0
	&
	&
	\\	-1
	&
	&
	&
	\\
	&
	&
	&	+1
	\\
	&
	&	0
	&
	\end{pmatrix}
	\equiv - \tau_{-+}^{z}
.\end{align}
\end{subequations}
Here, we defined $\tau_{s \bar{s}}^{\mu}$ ($s = \pm$ and $\mu = 0, x, y, z$).

To summarize these, we have
\begin{subequations}
\begin{align}
P_{s} \bm{\rho}^{\perp} \sigma^0
	& = \bm{\tau}^{\perp}_{s \bar{s}}
, \\
P_{s} \bm{\rho}^{\perp} \sigma^z
	& = s \tau^z_{s \bar{s}} \bm{\tau}^{\perp}_{s \bar{s}}
\notag \\
	& = - \zi s ( \hat{z} \times \bm{\tau}^{\perp}_{s \bar{s}} )
, \\
P_{s} \rho_{\mu = 0, 3} \sigma^x
	& = \tau^{\mu = 0, z}_{s \bar{s}}
, \\
P_{s} \rho_{\mu = 0, 3} \sigma^y
	& = - \zi s \tau^{\mu = z, 0}_{s \bar{s}}
.\end{align}
\label{eqs:P_eta-non-commutative}
\end{subequations}

In the main text, we denote $\tau_{ss}^{\mu}$ as $\tau^{\mu}$ for readability, and in this supplemental material we use the explicit expressions.

\section{\label{sec:nonadiabatic_components}Nonadiabatic components of spin and spin current}
Here, we show the nonadiabatic components of spin and spin current operators, which are terms mixed between the bonding and antibonding states.
In the rotated frame, the spin operator is given by
\begin{align}
s^{\alpha} (\bq)
	& = \sum_{\bk} \Psi_{\bk-\bq}^{\dagger} \rho_0 \sigma^{\alpha} \Psi_{\bk}^{}
.\end{align}
After the unitary transformation by $V_{\bk}$ and the projection operation, we have the adiabatic and nonadiabatic components of the spin:
\begin{subequations}
\begin{align}
s^{x} (\bq)
	& = \sum_{\bk, s} \psi_{\bk-\frac{\bq}{2}, s}^{\dagger} \left(
		s \sin \vartheta_{\bk} \hat{e}_{\varphi_{\bk}} \cdot \bm{\tau}_{s s}^{\perp}
	\right) \psi_{\bk+\frac{\bq}{2}, s}^{}
	+ \sum_{\bk, s} \psi_{\bk-\frac{\bq}{2}, s}^{\dagger} \left(
		- \cos \vartheta_{\bk} \tau_{s \bar{s}}^0
	\right) \psi_{\bk+\frac{\bq}{2}, \bar{s}}^{}
, \\
s^{y} (\bq)
	& = \sum_{\bk, s} \psi_{\bk-\frac{\bq}{2}, s}^{\dagger} \left(
		\sin \vartheta_{\bk} (\hat{z} \times \hat{e}_{\varphi_{\bk}}) \cdot \bm{\tau}_{s s}^{\perp}
	\right) \psi_{\bk+\frac{\bq}{2}, s}^{}
	+ \sum_{\bk, s} \psi_{\bk-\frac{\bq}{2}, s}^{\dagger} \left(
		\zi s \cos \vartheta_{\bk} \tau_{s \bar{s}}^z
	\right) \psi_{\bk+\frac{\bq}{2}, \bar{s}}^{}
, \\
s^{z} (\bq)
	& = \sum_{\bk, s} \psi_{\bk-\frac{\bq}{2}, s}^{\dagger} \left(
			s \tau_{s s}^z
		\right) \psi_{\bk+\frac{\bq}{2}, s}^{}
.\end{align}\label{eq:spins}
\end{subequations}
Note that $s^z$ does not have the nonadiabatic component in the lowest order of $\bq$.
By neglecting the nonadiabatic components, we obtain the Eq.~(8) with (9) in the main text:
\begin{align}
s^{\alpha} (\bq)
	& = \sum_{s = \pm} \sum_{\bk} \psi_{\bk-\frac{\bq}{2}, s}^{\dagger} \left(
			s \tau_{\bk, s}^{\alpha}
		\right) \psi_{\bk+\frac{\bq}{2}, s}^{}
\label{eq:spin}
,\end{align}
where $\bm{\tau}_{\bk, s} = (\tau_{\bk, s}^x, \tau_{\bk, s}^y, \tau_{\bk, s}^z)$ is given as
\begin{align}
\bm{\tau}_{\bk, s}
	& = \begin{pmatrix}
		\sin \vartheta_{\bk} \hat{e}_{\varphi_{\bk}} \cdot \bm{\tau}^{\perp}_{s s}
	\\	s \sin \vartheta_{\bk} \bigl( \hat{z} \times \hat{e}_{\varphi_{\bk}} \bigr) \cdot \bm{\tau}^{\perp}_{s s}
	\\	\tau^z_{s s}
	\end{pmatrix}
\label{eq:spin-vector-of-bonding_state}
\end{align}
(In the main text, we denote $\tau_{s s}^{\mu}$ as $\tau^{\mu}$ for readability.)
We here define the nonadiabatic components as $\bm{s}_{\mathrm{na}}^{\perp} = (s^x_{\mathrm{na}}, s^y_{\mathrm{na}}, 0)$;
\begin{align}
\bm{s}_{\mathrm{na}}^{\perp} (\bq)
	& = \sum_{\bk, s} \psi_{\bk-\frac{\bq}{2}, s}^{\dagger}
	\begin{pmatrix}
		- \cos \vartheta_{\bk} \tau_{s \bar{s}}^0
	\\	\zi s \cos \vartheta_{\bk} \tau_{s \bar{s}}^z
	\\	0
	\end{pmatrix}
	\psi_{\bk+\frac{\bq}{2}, \bar{s}}^{}
\label{eq:spin_na}
.\end{align}
This expression are used in Sec.~\ref{sec:nonadiabatic_torques}.

The spin current operator in the rotated frame is given by
\begin{align}
j_{\mathrm{s}, i}^{\alpha} (\bq)
	& = \sum_{\bk} \Psi_{\bk-\frac{\bq}{2}}^{\dagger}
		\left( \frac{1}{\hbar} \frac{\partial H_{\bk}}{\partial k_i} \sigma^{\alpha} \right)
	\Psi_{\bk+\frac{\bq}{2}}
.\end{align}
After the unitary transformation by $V_{\bk}$ and the projection operation, we have the spin current operator in the eigenstate frame;
\begin{align}
j_{\mathrm{s}, i}^{x} (\bq)
	& = \sum_{\bk, s} \psi_{\bk-\frac{\bq}{2} s}^{\dagger} \left\{
		\frac{s}{\hbar} \frac{\partial \epsilon_{\bk s}}{\partial k_i} \tau_{\bk, s}^x
		+ \frac{\mathcal{J}_{sd}}{\hbar} (\hat{z} \times \bm{A}_{\mathrm{k}, i}^{\perp}) \cdot \bm{\tau}_{s s}^{\perp}
	\right\} \psi_{\bk+\frac{\bq}{2} s}
\notag \\ & \hspace{1em}
	+ \sum_{\bk, s} \psi_{\bk-\frac{\bq}{2} s}^{\dagger} \left[
		- \frac{1}{\hbar} \frac{\partial \epsilon_{\bk s}}{\partial k_i} \cos \vartheta_{\bk} \tau_{s \bar{s}}^0
		- \frac{\zi s \Delta_{\bk} \sin \vartheta_{\bk}}{\hbar} \left\{
			\bm{A}_{\mathrm{k}, i}^{\perp} \cdot \hat{e}_{\varphi_{\bk}} \tau_{s \bar{s}}^z
			+ \zi ( \bm{A}_{\mathrm{k}, i}^{\perp} \times \hat{e}_{\varphi_{\bk}} )^z \tau_{s \bar{s}}^0
		\right\}
	\right] \psi_{\bk+\frac{\bq}{2} \bar{s}}
, \\
j_{\mathrm{s}, i}^{y} (\bq)
	& = \sum_{\bk, s} \psi_{\bk-\frac{\bq}{2} s}^{\dagger} \left(
		\frac{s}{\hbar} \frac{\partial \epsilon_{\bk s}}{\partial k_i} \tau_{\bk, s}^y
		- \frac{s \mathcal{J}_{sd}}{\hbar} \bm{A}_{\mathrm{k}, i}^{\perp} \cdot \bm{\tau}_{s s}^{\perp}
	\right) \psi_{\bk+\frac{\bq}{2} s}
\notag \\ & \hspace{1em}
	+ \sum_{\bk, s} \psi_{\bk-\frac{\bq}{2} s}^{\dagger} \left[
		\frac{\zi s}{\hbar} \frac{\partial \epsilon_{\bk s}}{\partial k_i} \cos \vartheta_{\bk} \tau_{s \bar{s}}^{z}
		- \frac{\Delta_{\bk} \sin \vartheta_{\bk}}{\hbar} \left\{
			\bm{A}_{\mathrm{k}, i}^{\perp} \cdot \hat{e}_{\varphi_{\bk}} \tau_{s \bar{s}}^0
			+ \zi ( \bm{A}_{\mathrm{k}, i}^{\perp} \times \hat{e}_{\varphi_{\bk}} )^z \tau_{s \bar{s}}^z
		\right\}
	\right] \psi_{\bk+\frac{\bq}{2} \bar{s}}
, \\
j_{\mathrm{s}, i}^{z} (\bq)
	& = \sum_{\bk, s} \psi_{\bk-\frac{\bq}{2} s}^{\dagger} \left(
		\frac{s}{\hbar} \frac{\partial \epsilon_{\bk s}}{\partial k_i} \tau_{s s}^z
	\right) \psi_{\bk+\frac{\bq}{2} s}^{}
	+ \sum_{\bk, s} \psi_{\bk-\frac{\bq}{2} s}^{\dagger} \left(
		\zi s \frac{\Delta_{\bk}}{\hbar} \bm{A}^{\perp}_{\mathrm{k}, i} \cdot \bm{\tau}_{s \bar{s}}^{\perp}
	\right) \psi_{\bk+\frac{\bq}{2} \bar{s}}^{}
,\end{align}
where $\tau^x_{\bk, s}$ and $\tau^y_{\bk, s}$ are given in Eq.~(\ref{eq:spin}).
By neglecting the nonadiabatic components and dividing into the ordinary spin current $\bm{j}_{\mathrm{s}0,i}$ and the anomalous spin current $\bm{j}_{\mathrm{sA}, i}$, we have Eq.~(10) and (11) in the main text:
\begin{align}
\bm{j}_{\mathrm{s0}, i} (\bq)
	& = \sum_{\bk, s} \psi_{\bk-\frac{\bq}{2}, s}^{\dagger} \left(
		\frac{s}{\hbar} \frac{\partial \epsilon_{\bk s}}{\partial k_i} \bm{\tau}_{\bk, s}
	\right) \psi_{\bk+\frac{\bq}{2}, s}
\label{eq:j_s0}
\end{align}
and
\begin{align}
\bm{j}_{\mathrm{sA}, i} (\bq)
	& = \frac{\mathcal{J}_{sd}}{\hbar} \sum_{\bk, s} \psi_{\bk-\frac{\bq}{2}, s}^{\dagger}
		\begin{pmatrix}
			(\hat{z} \times \bm{A}_{\mathrm{k}, i}^{\perp}) \cdot \bm{\tau}_{s s}^{\perp}
		\\	- s \bm{A}_{\mathrm{k}, i}^{\perp} \cdot \bm{\tau}_{s s}^{\perp}
		\\	0
		\end{pmatrix}
	\psi_{\bk+\frac{\bq}{2}, s}
\label{eq:j_sA}
.\end{align}
Here, we define the nonadiabatic components of the spin current as
\begin{align}
\bm{j}_{\mathrm{s}, i}^{\mathrm{na}} (\bq)
	& = \sum_{\bk, s} \psi_{\bk-\frac{\bq}{2} s}^{\dagger}
			\bm{J}_{i, \bk, s}
		\psi_{\bk+\frac{\bq}{2} \bar{s}}^{}	
		+ \hat{z} \sum_{\bk, s} \psi_{\bk-\frac{\bq}{2} s}^{\dagger} \left(
			\zi s \frac{\Delta_{\bk}}{\hbar} \bm{A}^{\perp}_{\mathrm{k}, i} \cdot \bm{\tau}_{s \bar{s}}^{\perp}
		\right)
		\psi_{\bk+\frac{\bq}{2} \bar{s}}^{}	
\end{align}
with
\begin{align}
\bm{J}_{i, \bk, s}
	& = \left( \frac{1}{\hbar} \frac{\partial \epsilon_{\bk s}}{\partial k_i} \cos \vartheta_{\bk} \right)
	\begin{pmatrix}
		- \tau_{s \bar{s}}^0
	\\	\zi s \tau_{s \bar{s}}^{z}
	\\	0
	\end{pmatrix}
	+ \frac{\Delta_{\bk} \sin \vartheta_{\bk}}{\hbar} \left\{
			\left( \bm{A}_{\mathrm{k}, i}^{\perp} \cdot \hat{e}_{\varphi_{\bk}} \right)
		\begin{pmatrix}
			- \zi s \tau_{s \bar{s}}^z
		\\	- \tau_{s \bar{s}}^0
		\\	0
		\end{pmatrix}
			+ ( \bm{A}_{\mathrm{k}, i}^{\perp} \times \hat{e}_{\varphi_{\bk}} )^z
		\begin{pmatrix}
			s \tau_{s \bar{s}}^0
		\\	- \zi \tau_{s \bar{s}}^z
		\\	0
		\end{pmatrix}
	\right\}
\label{eq:J_i_k_s}
\end{align}
We use these expressions in Sec.~\ref{sec:nonadiabatic_torques}.

\section{\label{sec:spin_torques}Spin torques in eigenstate frame}
In this section, we show the adiabatic and nonadiabatic components of the spin torques.
Since the spin torques are given as
\begin{align}
(\mathcal{R}^{-1} \bm{\tau}_m)_{\bq}
	& = \frac{\mathcal{J}_{sd}}{\hbar} \hat{z} \times \langle \bm{\Sigma} (\bq, t) \rangle_{\mathrm{neq}}
, \\
(\mathcal{R}^{-1} \bm{\tau}_n)_{\bq}
	& = \frac{J}{\hbar} \hat{z} \times \langle \bm{s} (\bq, t) \rangle_{\mathrm{neq}}
,\end{align}
we need to know the adiabatic and nonadiabatic components of $\bm{\Sigma} (\bq) = \sum_{\bk} \Psi_{\bk-\frac{\bq}{2}} \rho_3 \sigma^{\alpha} \Psi_{\bk+\frac{\bq}{2}}$ and $\bm{s} (\bq)$.
The expression for $\bm{s} (\bq)$ is already obtained in Eqs.~(\ref{eq:spins}).
Hence, we calculate $\bm{\Sigma} (\bq)$ in the same manner as calculated $\bm{s} (\bq)$, and then we have
\begin{align}
\Sigma^x (\bq)
	& = \sum_{\bk} \sum_{s} \psi_{\bk-\frac{\bq}{2} s}^{\dagger} \left(
		\frac{\zi s q_j}{2} \bm{A}_{\mathrm{k}, j}^{\perp} \cdot \bm{\tau}_{s s}^{\perp}
	\right) \psi_{\bk+\frac{\bq}{2} s}^{}
	+ \sum_{\bk} \sum_{s} \psi_{\bk-\frac{\bq}{2} s}^{\dagger} \left(
		- \tau_{s \bar{s}}^z
		- \frac{\zi q_j}{2} A_{\mathrm{k}, j}^z \tau_{s \bar{s}}^0
	\right) \psi_{\bk+\frac{\bq}{2} \bar{s}}^{}
, \\
\Sigma^y (\bq)
	& = \sum_{\bk} \sum_{s} \psi_{\bk-\frac{\bq}{2} s}^{\dagger} \left(
		\frac{\zi q_j}{2} (\hat{z} \times \bm{A}_{\mathrm{k}, j}^{\perp}) \cdot \bm{\tau}_{s s}^{\perp}
	\right) \psi_{\bk+\frac{\bq}{2} s}^{}
	+ \sum_{\bk} \sum_{s} \psi_{\bk-\frac{\bq}{2} s}^{\dagger} \left(
		\zi s \tau_{s \bar{s}}^0
		- \frac{s q_j}{2} A_{\mathrm{k}, j}^z \tau_{s \bar{s}}^z
	\right) \psi_{\bk+\frac{\bq}{2} \bar{s}}^{}
, \\
\Sigma^z (\bq)
	& = \sum_{\bk} \sum_{s} \psi_{\bk-\frac{\bq}{2} s}^{\dagger} \left(
		s \cos \vartheta_{\bk} \tau_{s s}^0
		- \frac{\zi s q_j}{2} A_{\mathrm{k}, j}^z \tau_{s s}^z
	\right) \psi_{\bk+\frac{\bq}{2} s}^{}
	+ \sum_{\bk} \sum_{s} \psi_{\bk-\frac{\bq}{2} s}^{\dagger} \left(
		\sin \vartheta_{\bk} \hat{e}_{\varphi_{\bk}} \cdot \bm{\tau}_{s \bar{s}}^{\perp}
	\right) \psi_{\bk+\frac{\bq}{2} \bar{s}}^{}
,\end{align}
where we keep the first order with respect to $\bq$.
By using the vector form, in the adiabatic regime, we find
\begin{align}
\bm{\Sigma} (\bq)
	& = \hat{z} \sum_{\bk} \sum_{s} \psi_{\bk-\frac{\bq}{2} s}^{\dagger} \left(
		s \cos \vartheta_{\bk} \tau_{s s}^0
		- \frac{\zi q_j}{2} A_{\mathrm{k}, j}^z \tau_{s s}^z
	\right) \psi_{\bk+\frac{\bq}{2} s}^{}
	+ \frac{\zi q_j}{2} \sum_{\bk} \sum_{s} \psi_{\bk-\frac{\bq}{2} s}^{\dagger}
		\begin{pmatrix}
			s \bm{A}_{\mathrm{k}, j}^{\perp} \cdot \bm{\tau}_{s s}^{\perp}
		\\	(\hat{z} \times \bm{A}_{\mathrm{k}, j}^{\perp}) \cdot \bm{\tau}_{s s}^{\perp}
		\\	0
		\end{pmatrix}
	\psi_{\bk+\frac{\bq}{2} s}^{}
\notag \\
	& = \hat{z} \sum_{\bk} \sum_{s} \psi_{\bk-\frac{\bq}{2} s}^{\dagger} \left(
		s \cos \vartheta_{\bk} \tau_{s s}^0
		- \frac{\zi q_j}{2} A_{\mathrm{k}, j}^z \tau_{s s}^z
	\right) \psi_{\bk+\frac{\bq}{2} s}^{}
	- \frac{\zi \hbar q_j}{2 \mathcal{J}_{sd}} (\hat{z} \times \bm{j}_{\mathrm{sA}, j} (\bq))
.\end{align}
Hence, we have the Eq.~(14a) in the main text.

For the nonadiabatic component, we find $\bm{\Sigma}_{\mathrm{na}}^{\perp} = (\Sigma_{\mathrm{na}}^x, \Sigma_{\mathrm{na}}^y, 0)$ with
\begin{align}
\bm{\Sigma}_{\mathrm{na}}^{\perp} (\bq)
	& = \sum_{\bk} \sum_{s} \psi_{\bk-\frac{\bq}{2} s}^{\dagger} \left\{
	\begin{pmatrix}
		- \tau_{s \bar{s}}^z
	\\[1ex]	\zi s \tau_{s \bar{s}}^0
	\\[1ex]	0
	\end{pmatrix}
	+ \mathcal{O} (q)
	\right\}
	\psi_{\bk+\frac{\bq}{2} \bar{s}}^{}
\label{eq:Sigma_na}
.\end{align}
We use this expression in Sec.~\ref{sec:nonadiabatic_torques}.

\section{\label{sec:current}Electric current operator in eigenstate frame}
In order to calculate the current-induced spin-transfer torques based on the linear response theory, we derive the charge current operator in the eigenstate frame.
The electromagnetic coupling in the rotated frame is given as
\begin{align}
\mathcal{H}_{\mathrm{em}}
	& = - e \int_{\omega} \sum_{\bq} \bm{j} (-\bq) \cdot \bm{A}^{\mathrm{em}} (\bq, \omega)
\end{align}
with the electric current operator $\bm{j} = \bm{j}_{\mathrm{p}} + \bm{j}_{\mathrm{A}}$, where $\bm{j}_{\mathrm{p}}$ is the paramagnetic current given by $\bm{j}_{\mathrm{p}} = (j_{\mathrm{p},x}, j_{\mathrm{p},y}, j_{\mathrm{p},z})$ with
\begin{align}
j_{\mathrm{p}, i} (\bq)
	& = \sum_{\bk} \Psi_{\bk-\frac{\bq}{2}}^{\dagger}
	\left( \frac{1}{\hbar} \frac{\partial H_{\bk}}{\partial k_i} \right)
	\Psi_{\bk+\frac{\bq}{2}}
\label{eq:j_p_rotated}
,\end{align}
and $\bm{j}_{\mathrm{A}}$ is the anomalous current given by $\bm{j}_{\mathrm{A}} = (j_{\mathrm{A},x}, j_{\mathrm{A},y}, j_{\mathrm{A},z})$ with
\begin{align}
j_{\mathrm{A}, i} (\bq) 
 	& = \frac{\hbar}{2} \int_{\omega'} \sum_{\bk, \bq'} A_{\mathrm{r}, j}^{\alpha} (\bq + \bq', \omega')
	\Psi_{\bk+\frac{\bq'}{2}}^{\dagger}
		\left( \frac{1}{\hbar^2} \frac{\partial^2 H_{\bk}}{\partial k_i \partial k_j} \right) \sigma^{\alpha}
	\Psi_{\bk-\frac{\bq'}{2}}
\label{eq:j_A_rotated}
.\end{align}
In the eigenstate frame, Eqs.~(\ref{eq:j_p_rotated}) and (\ref{eq:j_A_rotated}) are obtained as
\begin{align}
j_{\mathrm{p}, i} (\bq)
	& = \sum_{\bk,s,s'} \psi_{\bk-\frac{\bq}{2} s}^{\dagger} \left(
		\frac{1}{\hbar} \frac{\partial \epsilon_{\bk s}}{\partial k_i} \tau_{s s}^{0}
	\right) \psi_{\bk+\frac{\bq}{2} s}^{}
	+ \sum_{\bk,s,s'} \psi_{\bk-\frac{\bq}{2} s}^{\dagger} \left\{
		- \frac{\Delta_{\bk}}{\hbar} ( \hat{z} \times \bm{A}_{\mathrm{k}, i}^{\perp} ) \cdot \bm{\tau}_{s \bar{s}}^{\perp}
	\right\} \psi_{\bk+\frac{\bq}{2} \bar{s}}^{}
\label{eq:j_p}
,\end{align}
and $j_{\mathrm{A}, i} = j^{(1)}_{\mathrm{A}, i} + j^{(2)}_{\mathrm{A}, i}$, 
\begin{align}
j^{(1)}_{\mathrm{A}, i} (\bq)
 	& = \frac{\hbar}{2} \int_{\omega'} \sum_{\bq'} A_{\mathrm{r}, j}^{\alpha} (\bq + \bq', \omega')
		\mathscr{N}_{1, i j}^{\alpha} (-\bq')
	+ j_{\mathrm{A},i}^{(1), \mathrm{na}} (\bq)
\label{eq:j_A1}
, \\
j^{(2)}_{\mathrm{A}, i} (\bq)
 	& = \frac{\hbar}{2} \int_{\omega'} \sum_{\bq'} A_{\mathrm{r}, j}^{\alpha} (\bq + \bq', \omega')
		\mathscr{N}_{2, i j}^{\alpha} (-\bq')
	+ j_{\mathrm{A},i}^{(2), \mathrm{na}} (\bq)
\label{eq:j_A2}
\end{align}
with
\begin{align}
\mathscr{N}_{1, i j}^{\alpha} (\bq)
	& = \sum_{\bk, s} \psi_{\bk-\frac{\bq}{2}}^{\dagger} \left[
		\left(
			\frac{s}{\hbar^2} \frac{\partial^2 \epsilon_{\bk s}}{\partial k_i \partial k_j}
			+ \frac{\Delta_{\bk}}{\hbar^2} \bm{A}_{\mathrm{k}, i}^{\perp} \cdot \bm{A}_{\mathrm{k}, j}^{\perp}
		\right) \tau_{\bk, s}^{\alpha}
	\right] \psi_{\bk+\frac{\bq}{2} s}^{}
, \\
\mathscr{N}_{2, i j}^{ \alpha} (\bq)
	& = \frac{1}{\hbar^2} \sum_{\bk, s} \psi_{\bk-\frac{\bq}{2}}^{\dagger}
	\begin{pmatrix}
		\cos \vartheta_{\bk} \bm{P}_{i j} \cdot \bm{\tau}_{s s}^{\perp}
	\\	 s \cos \vartheta_{\bk} (\hat{z} \times \bm{P}_{i j}) \cdot \bm{\tau}_{s s}^{\perp}
	\\	0
	\end{pmatrix}^{\alpha}
	\psi_{\bk+\frac{\bq}{2} s}^{}
,\end{align}
where $\bm{P}_{ij} = \bm{P}_{ji}$ is defined by
\begin{align}
\bm{P}_{i j}
	& = \hat{z} \times \left(
		\frac{\partial \Delta_{\bk}}{\partial k_j} \bm{A}_{\mathrm{k}, i}^{\perp}
		+ \frac{\partial \Delta_{\bk}}{\partial k_i} \bm{A}_{\mathrm{k}, j}^{\perp}
		+ \frac{\Delta_{\bk}}{2} \frac{\partial \bm{A}_{\mathrm{k}, i}^{\perp}}{\partial k_j}
		+ \frac{\Delta_{\bk}}{2} \frac{\partial \bm{A}_{\mathrm{k}, j}^{\perp}}{\partial k_i}
	\right)
	+ \frac{\Delta_{\bk}}{2} \left(
		A_{\mathrm{k}, j}^z \bm{A}_{\mathrm{k}, i}^{\perp}
		+ A_{\mathrm{k}, i}^z \bm{A}_{\mathrm{k}, j}^{\perp}
	\right)
.\end{align}
Here, $j_{\mathrm{A},i}^{(1), \mathrm{na}} (\bq)$ and $j_{\mathrm{A},i}^{(2), \mathrm{na}} (\bq)$ are the nonadiabatic components of the anomalous current, which are found to be negligible for the calculation in the nonadiabatic spin-transfer torques.

\section{\label{sec:adiabatic_torques}Adiabatic spin-transfer torques}
Now, we show the calculation of the adiabatic spin-transfer torques.
As shown in the main text, the adiabatic spin-transfer torques are obtained as the linear response of the anomalous spin current and the spin to the vector potential.
\begin{align}
\langle j_{\mathrm{sA}, i}^{\alpha} (\bq, \omega) \rangle_{\mathrm{ne}}
	& = X_{i j}^{\alpha} (\bq, \omega) A_{j}^{\mathrm{em}} (\omega)
, \\
\langle s^{\alpha} (\bq, \omega) \rangle_{\mathrm{ne}}
	& = Y_{j}^{\alpha} (\bq, \omega) A_{j}^{\mathrm{em}} (\omega)
,\end{align}
where the linear response coefficients $X_{i j}^{\alpha} (\bq, \omega)$ and $Y_{j}^{\alpha} (\bq, \omega)$ are calculated from the corresponding correlation functions,
\begin{align}
\mathscr{X}_{i j}^{\alpha} (\bq, \zi \omega_{\lambda})
	& = \frac{- e}{V} \int_0^{\beta} \dd{\tau} e^{\zi \omega_{\lambda} \tau} \left\langle
		\mathrm{T}_{\tau} \left\{ j_{\mathrm{sA}, i}^{\alpha} (\bq, \tau) j_{j} (0, 0) \right\}
	\right\rangle
, \\
\mathscr{Y}_{i}^{\alpha} (\bq, \zi \omega_{\lambda})
	& = \frac{- e}{V} \int_0^{\beta} \dd{\tau} e^{\zi \omega_{\lambda} \tau} \left\langle
		\mathrm{T}_{\tau} \left\{ s^{\alpha} (\bq, \tau) j_{j} (0, 0) \right\}
	\right\rangle
\end{align}
by taking the analytic continuation $\zi \omega_{\lambda} \to \hbar \omega + \zi 0$;
\begin{align}
X_{i j}^{\alpha} (\bq, \omega)
	& = \mathscr{X}_{i j}^{\alpha} (\bq, \hbar \omega + \zi 0)
, \\
Y_{j}^{\alpha} (\bq, \omega)
	& = \mathscr{Y}_{j}^{\alpha} (\bq, \hbar \omega + \zi 0)
.\end{align}

\subsection{Linear response coefficient for adiabatic component of anomalous spin current}
First, we calculate the coefficient $\mathscr{X}_{i j}^{\alpha} (\bq, \zi \omega_{\lambda})$.
We divide $\mathscr{X}_{i j}^{\alpha} (\bq, \zi \omega_{\lambda})$ into the three terms,
\begin{align}
\mathscr{X}_{i j}^{\alpha} (\bq, \zi \omega_{\lambda})
	& = \mathscr{X}_{i j}^{(1) \alpha}
		+ \mathscr{X}_{i j}^{(2) \alpha}
		+ \mathscr{X}_{i j}^{(3) \alpha}
,\end{align}
where
\begin{align}
\mathscr{X}_{i j}^{(1) \alpha}
	& = \frac{-e}{V} \int_0^{\beta} \dd{\tau} e^{\zi \omega_{\lambda} \tau} \left\langle
		\mathrm{T}_{\tau} \left\{ j_{\mathrm{sA}, i}^{\alpha} (\bq, \tau) j_{\mathrm{p}, j} (0, 0) \right\}
	\right\rangle
, \\
\mathscr{X}_{i j}^{(2) \alpha}
	& = \frac{-e}{V} \int_0^{\beta} \dd{\tau} e^{\zi \omega_{\lambda} \tau} \left\langle
		\mathrm{T}_{\tau} \left\{ j_{\mathrm{sA}, i}^{\alpha} (\bq, \tau) j_{\mathrm{A}, j}^{(1)} (0, 0) \right\}
	\right\rangle
, \\
\mathscr{X}_{i j}^{(3) \alpha}
	& = \frac{-e}{V} \int_0^{\beta} \dd{\tau} e^{\zi \omega_{\lambda} \tau} \left\langle
		\mathrm{T}_{\tau} \left\{ j_{\mathrm{sA}, i}^{\alpha} (\bq, \tau) j_{\mathrm{A}, j}^{(2)} (0, 0) \right\}
	\right\rangle
.\end{align}
Here, $j_{\mathrm{p}, j}$ is given by Eq.~(\ref{eq:j_p}), and $j_{\mathrm{A}, j}^{(1)}$ and $j_{\mathrm{A}, j}^{(2)}$ are given by Eqs.~(\ref{eq:j_A1}) and (\ref{eq:j_A2}).
However, $\mathscr{X}_{i j}^{(2) \alpha}$ and $\mathscr{X}_{i j}^{(3) \alpha}$ are higher order contributions with respect to $\bq$, since the anomalous current is proportional to the spin gauge field, which is first order of $\bq$.
Hence, we neglect $\mathscr{X}_{i j}^{(2) \alpha}$ and $\mathscr{X}_{i j}^{(3) \alpha}$.

Rewriting $\mathscr{X}_{i j}^{(1) \alpha}$ by means of the Green function and expanding the first order of $\mathcal{H}_f$, we have
\begin{align}
\mathscr{X}_{i j}^{(1) \alpha}
	& = - \frac{M}{N} \frac{e \mathcal{J}_{sd}^2}{\hbar} (\mathcal{R}^{-1} \bm{m})^{\beta}_{\bq}
		\frac{1}{\beta V} \sum_n \sum_{\bk, s} \frac{1}{\hbar} \frac{\partial \epsilon_{\bk s}}{\partial k_j}
		\tr \left[
			C^{\alpha \gamma}_{\bk, i, s} \tau^{\gamma}_{s s} \tau_{\bk, s}^{\beta}
		\right]
		\left\{
			(g_{\bk,s}^{+})^2 g_{\bk,s}
			+ g_{\bk,s}^{+} (g_{\bk,s})^2
		\right\}
,\end{align}
where $g_{\bk,s}^{+} = g_{\bk,s} (\zi \epsilon_n + \zi \omega_{\lambda})$, $g_{\bk,s} = g_{\bk,s} (\zi \epsilon_n)$, and
\begin{align}
\begin{pmatrix}
	(\hat{z} \times \bm{A}_{\mathrm{k}, i}^{\perp}) \cdot \bm{\tau}_{s s}^{\perp}
\\	- s \bm{A}_{\mathrm{k}, i}^{\perp} \cdot \bm{\tau}_{s s}^{\perp}
\\	0
\end{pmatrix}
	& = \begin{pmatrix}
		- A_{\mathrm{k}, i}^y	
	&	A_{\mathrm{k}, i}^x 	
	&	0
	\\	- s A_{\mathrm{k}, i}^x	
	&	- s A_{\mathrm{k}, i}^y	
	&	0
	\\	0
	&	0
	&	0
	\end{pmatrix}^{\alpha \beta}
	\begin{pmatrix}
		\tau^x_{s s}
	\\	\tau^y_{s s}
	\\	\tau^z_{s s}
	\end{pmatrix}^{\beta}
	= \mathcal{C}_{\bk s, i} \bm{\tau}_{s s}^{\beta}
.\end{align}
Here, we keep the $\bq$ dependence only of $(\mathcal{R}^{-1} \bm{m})^{\beta}_{\bq}$ and set $\bq = 0$ for the other parts, since the $\bq$ contributions from the other parts  are higher orders.
Here, we calculate the trace as
\begin{align}
\tr \left[ C^{\alpha \gamma}_{\bk, i, s} \tau^{\gamma}_{s s} \tau_{\bk, s}^{\beta} \right]
	& = 2 \sin \vartheta_{\bk} (\bm{A}^{\perp}_{\mathrm{k}, i} \times \hat{e}_{\varphi_{\bk}})^z \delta^{\alpha \beta}
		+ 2 s \sin \vartheta_{\bk} (\bm{A}^{\perp}_{\mathrm{k}, i} \cdot \hat{e}_{\varphi_{\bk}}) \epsilon^{\alpha \beta z}
\notag \\
	& = 2 \left( \hat{z} \cdot \frac{\partial \bm{x}_{\bk}}{\partial k_i} \right) \delta^{\alpha \beta}
		- 2 s \left\{ \hat{z} \cdot \left( \bm{x}_{\bk} \times \frac{\partial \bm{x}_{\bk}}{\partial k_i} \right) \right\} \epsilon^{\alpha \beta z}
,\end{align}
where
\begin{align}
\bm{x}_{\bk}
	& = \begin{pmatrix}
		\sin \vartheta_{\bk} \cos \varphi_{\bk}
	\\	\sin \vartheta_{\bk} \sin \varphi_{\bk}
	\\	\cos \vartheta_{\bk}
	\end{pmatrix}
,\end{align}
and we use the relation
\begin{align}
\sin \vartheta_{\bk} (\bm{A}^{\perp}_{\mathrm{k}, i} \times \hat{e}_{\varphi_{\bk}})^z
	& = \left( \mathcal{R}_{\mathrm{k}} \hat{z} \right) \cdot \left[ \left( \mathcal{R}_{\mathrm{k}} \bm{A}^{\perp}_{\mathrm{k}, i} \right) \times \mathcal{R}_{\mathrm{k}} ( \sin \vartheta_{\bk} \hat{e}_{\varphi_{\bk}} ) \right]
	= \hat{z} \cdot \frac{\partial \bm{x}_{\bk}}{\partial k_i}
\label{eq:A_x_e}
, \\
\sin \vartheta_{\bk} (\bm{A}^{\perp}_{\mathrm{k}, i} \cdot \hat{e}_{\varphi_{\bk}}) 
	& = \left( \mathcal{R}_{\mathrm{k}} \bm{A}^{\perp}_{\mathrm{k}, i} \right) \cdot \mathcal{R}_{\mathrm{k}} ( \sin \vartheta_{\bk} \hat{e}_{\varphi_{\bk}} )
	= - \hat{z} \cdot \left( \bm{x}_{\bk} \times \frac{\partial \bm{x}_{\bk}}{\partial k_i} \right)
\label{eq:A_dot_e}
\end{align}
with $\mathcal{R}_{\mathrm{k}}^{\alpha \beta} = 2 \chi_{\bk}^{\alpha} \chi_{\bk}^{\beta} - \delta^{\alpha \beta}$, where $\chi^{\alpha} _{\bk}$ is defined by Eq.~(\ref{eq:chi}).
By using
\begin{align}
\frac{\partial \epsilon_{\bk s}}{\partial k_j} \left\{ (g_{\bk,s}^{+})^2 g_{\bk,s} + g_{\bk,s}^{+} (g_{\bk,s})^2 \right\}
	& = \frac{\partial}{\partial k_j} \left( g^{+}_{\bk,s} g_{\bk,s} \right)
,\end{align}
and integrating by parts, we have
\begin{align}
\mathscr{X}_{i j}^{(1) \alpha}
	& = \frac{M}{N} \Phi_{ij} (\zi \omega_{\lambda}) (\mathcal{R}^{-1} \bm{m})^{\alpha}_{\bq}
		- \frac{M}{N} \Phi'_{ij} (\zi \omega_{\lambda}) \left\{ (\mathcal{R}^{-1} \bm{m})_{\bq} \times \hat{z} \right\}^{\alpha}
\end{align}
with
\begin{align}
\Phi_{ij} (\zi \omega_{\lambda})
	& = \frac{2 e \mathcal{J}_{sd}^2}{\hbar^2} \frac{1}{\beta V} \sum_n \sum_{\bk, s}
		\frac{\partial^2 \cos \vartheta_{\bk}}{\partial k_i \partial k_j}
		g^{+}_{\bk,s} g_{\bk,s}
, \\
\Phi'_{ij} (\zi \omega_{\lambda})
	& = - \frac{2 e \mathcal{J}_{sd}^2}{\hbar^2} \frac{1}{\beta V} \sum_n \sum_{\bk, s} s
		\left\{ \hat{z} \cdot \frac{\partial}{\partial k_j} \left( \bm{x}_{\bk} \times \frac{\partial \bm{x}_{\bk}}{\partial k_i} \right) \right\}
		g^{+}_{\bk,s} g_{\bk,s}
.\end{align}
Then, we take the analytic continuation, so that we obtain
\begin{align}
\Phi_{ij} (\hbar \omega + \zi 0)
	& = - \zi \omega \frac{2 \mathcal{P}_m}{e} \sigma_{\mathrm{c}} \delta_{i,j}
, \\
\Phi'_{ij} (\hbar \omega + \zi 0)
	& = \zi \omega \frac{2 e \mathcal{J}_{sd}^2}{\hbar^2} \frac{1}{V} \sum_{\bk, s} s
		\left\{ \hat{z} \cdot \frac{\partial}{\partial k_j} \left( \bm{x}_{\bk} \times \frac{\partial \bm{x}_{\bk}}{\partial k_i} \right) \right\}
		\tau \delta (\mu - \epsilon_{\bk s})
,\end{align}
where $\tau$ is the lifetime of the bonding/antibonding state, $\sigma_{\mathrm{c}}$ is the conductivity given by
\begin{align}
\sigma_{\mathrm{c}}
	& = \frac{e^2}{V} \sum_{\bk,s} \left( \frac{1}{\hbar} \frac{\partial \epsilon_{\bk s}}{\partial k_i} \right)^2 \tau \delta (\mu - \epsilon_{\bk s})
,\end{align}
and the nondimensional coefficient $\mathcal{P}_m$ is given as
\begin{align}
\mathcal{P}_m
	& = \frac{e^2}{\sigma_{\mathrm{c}}} \frac{\mathcal{J}_{sd}^2}{\hbar^2} \frac{1}{V} \sum_{\bk, s}
		\left[
			\frac{2 \mathcal{J}_{sd}}{\Delta_{\bk}^3} \left( \frac{\partial \Delta_{\bk}}{\partial k_i} \right)^2
			- \frac{\mathcal{J}_{sd}}{\Delta_{\bk}^2} \frac{\partial^2 \Delta_{\bk}}{\partial^2 k_i}
		\right]
		\tau \delta (\mu - \epsilon_{\bk s})
\label{eq:P_m}
.\end{align}
For the case of $T_{\bk} = 0$, we have the simple form
\begin{align}
\mathcal{P}_m
	& \simeq 2 \mathcal{J}_{sd}^3 / |\mu|^3
, \qquad
( T_{\bk} = 0 )
.\end{align}

For $\Phi'_{ij}$, we calculate it in two cases; the case with site inversion symmetry and the honeycomb lattice for the simplest case with no site inversion symmetry.
For the case with site inversion symmetry,
\begin{align}
\bm{x}_{\bk}
	= \begin{pmatrix}
		\sin \vartheta_{\bk}
	\\	0
	\\	\cos \vartheta_{\bk}
	\end{pmatrix}
, \qquad
\hat{z} \cdot \frac{\partial}{\partial k_j} \left( \bm{x}_{\bk} \times \frac{\partial \bm{x}_{\bk}}{\partial k_i} \right)
	= 0
\label{eq:Phi'_with_SIS}
,\end{align}
so that $\Phi'_{ij} = 0$ in this case.
For the honeycomb lattice, setting $\eta = \pm$ as the valley degree of freedom,
\begin{align}
\bm{x}_{\bk}
	= \frac{1}{\Delta_{\bk}} \begin{pmatrix}
		\eta v k_x
	\\	v k_y
	\\	\mathcal{J}_{sd}
	\end{pmatrix}
, \qquad
\hat{z} \cdot \frac{\partial}{\partial k_j} \left( \bm{x}_{\bk} \times \frac{\partial \bm{x}_{\bk}}{\partial k_i} \right)
	& = \eta \frac{\partial}{\partial k_j} \left( \frac{v^2}{\Delta_{\bk}} ( k_x \delta_{i,y} - k_y \delta_{i,x} ) \right)
\label{eq:Phi'_without_SIS}
,\end{align}
and by summing $\eta = \pm$, we have $\Phi'_{ij} = 0$.

From above, we have
\begin{align}
\langle j_{\mathrm{sA}, i}^{\alpha} (\bq, \omega) \rangle_{\mathrm{ne}}
	& = \frac{M}{N} \frac{\mathcal{P}_m}{e} j_{\mathrm{c},i} (\mathcal{R}^{-1} \bm{m})^{\alpha}_{\bq}
,\end{align}
which leads to the result
\begin{align}
\bm{\tau} (\br, t)
	& = \frac{M}{N} \left( \frac{\mathcal{P}_m}{e} \bm{j}_{\mathrm{c}} \cdot \bm{\nabla} \right) \bm{m} (\br, t)
,\end{align}
where $\mathcal{P}_m$ is given by Eq.~(\ref{eq:P_m}).

\subsection{Linear response coefficient for adiabatic component of spin}
Then, we evaluate the linear response coefficient for the adiabatic component of spin $Y_{j}^{\alpha} (\bq, \omega)$.
In the similar way of the calculation $\mathscr{X}_{i j}^{\alpha} (\bq, \zi \omega_{\lambda})$, we divide $\mathscr{Y}_{j}^{\alpha} (\bq, \zi \omega_{\lambda})$ into three parts,
\begin{align}
\mathscr{Y}_{j}^{\alpha} (\bq, \zi \omega_{\lambda})
	& = \mathscr{Y}_{j}^{(1) \alpha}
		+ \mathscr{Y}_{j}^{(2) \alpha}
		+ \mathscr{Y}_{j}^{(3) \alpha}
\end{align}
with
\begin{align}
\mathscr{Y}_{i}^{(1) \alpha}
	& = \frac{-e}{V} \int_0^{\beta} \dd{\tau} e^{\zi \omega_{\lambda} \tau} \left\langle
		\mathrm{T}_{\tau} \left\{ \tilde{s}^{\alpha} (\bq, \tau) \tilde{j}_{\mathrm{p}, i} (0, 0) \right\}
	\right\rangle
, \\
\mathscr{Y}_{i}^{(2) \alpha}
	& = \frac{-e}{V} \int_0^{\beta} \dd{\tau} e^{\zi \omega_{\lambda} \tau} \left\langle
		\mathrm{T}_{\tau} \left\{ \tilde{s}^{\alpha} (\bq, \tau) \tilde{j}^{(1)}_{\mathrm{A}, i} (0, 0) \right\}
	\right\rangle
, \\
\mathscr{Y}_{i}^{(3) \alpha}
	& = \frac{-e}{V} \int_0^{\beta} \dd{\tau} e^{\zi \omega_{\lambda} \tau} \left\langle
		\mathrm{T}_{\tau} \left\{ \tilde{s}^{\alpha} (\bq, \tau) \tilde{j}^{(2)}_{\mathrm{A}, i} (0, 0) \right\}
	\right\rangle
.\end{align}
Rewriting them by means of the Green function and expanding the first order with respect to the spin gauge field, we have
\begin{align}
\mathscr{Y}_{i}^{(1) \alpha}
	& = \frac{e \hbar}{2 \beta V} A^{\beta}_{\mathrm{r}, j} (\bq) \sum_{n} \sum_{\bk, s}
		\frac{s}{\hbar} \frac{\partial \epsilon_{\bk s}}{\partial k_i}
		\mathrm{tr}\! \left[
			\tau^{\alpha}_{\bk s}
			\left(
				\frac{s}{\hbar} \frac{\partial \epsilon_{\bk s}}{\partial k_j} \tau_{\bk s}^{\beta}
				+ \frac{\mathcal{J}_{sd}}{\hbar} \mathcal{C}^{\beta \delta}_{\bk s, j} \tau^{\delta}_{s s}
			\right)
		\right]
		\left( (g^{+}_{\bk s})^2 g^{}_{\bk s} + g^{+}_{\bk s} (g^{}_{\bk s})^2 \right)
, \\
\mathscr{Y}_{i}^{(2) \alpha}
	& = \frac{e \hbar}{2 \beta V} A^{\beta}_{\mathrm{r}, j} (\bq) \sum_{n} \sum_{\bk, s}
		s \left(
			s \frac{1}{\hbar^2} \frac{\partial^2 \epsilon_{\bk s}}{\partial k_i \partial k_j}
			+ \frac{\Delta_{\bk}}{\hbar^2} \bm{A}_{\mathrm{k}, i}^{\perp} \cdot \bm{A}_{\mathrm{k}, j}^{\perp}
		\right)
		\mathrm{tr}\! \left[ \tau_{\bk s}^{\alpha} \tau_{\bk s}^{\beta} \right]
		g^{+}_{\bk s} g^{}_{\bk s}
, \\
\mathscr{Y}_{i}^{(3) \alpha}
	& = \frac{e \hbar}{2 \beta V} A^{\beta}_{\mathrm{r}, j} (\bq) \sum_{n} \sum_{\bk, s}
		\frac{1}{\hbar^2} \cos \vartheta_{\bk}
		\mathrm{tr}\! \left[ \tau_{\bk s}^{\alpha} \mathcal{D}_{i j, \bk s}^{\beta \delta} \tau^{\delta}_{s s} \right]
		g^{+}_{\bk s} g^{}_{\bk s}
,\end{align}
where 
\begin{align}
\begin{pmatrix}
	\bm{P}_{i j} \cdot \bm{\tau}_{s s}^{\perp}
\\	 s (\hat{z} \times \bm{P}_{i j}) \cdot \bm{\tau}_{s s}^{\perp}
\\	0
\end{pmatrix}
	& = \begin{pmatrix}
		P_{i j}^x
	&	P_{i j}^y
	&	0
	\\	- s P_{i j}^y
	&	s P_{i j}^x
	&	0
	\\	0
	&	0
	&	0
	\end{pmatrix}
	\begin{pmatrix}
		\tau_{s s}^x
	\\	\tau_{s s}^y
	\\	\tau_{s s}^z
	\end{pmatrix}
	= \mathcal{D}_{i j, \bk s} \bm{\tau}_{s s}
.\end{align}
Calculating $\tr[\cdots]$, we obtain
\begin{align}
\mathrm{tr}\! \left[ \tau^{\alpha}_{\bk s} \tau_{\bk s}^{\beta} \right]
	& = 2 \sin^2 \vartheta_{\bk} \delta^{\alpha \beta}
, \\
\mathrm{tr}\! \left[ \tau^{\alpha}_{\bk s} \mathcal{C}^{\beta \delta}_{\bk s, j} \tau^{\delta}_{s s} \right]
	& = 2 \sin \vartheta_{\bk} (\bm{A}^{\perp}_{\mathrm{k}, j} \times \hat{e}_{\varphi_{\bk}})^z \delta^{\alpha \beta}
		- 2 s \sin \vartheta_{\bk} (\bm{A}^{\perp}_{\mathrm{k}, j} \cdot \hat{e}_{\varphi_{\bk}}) \epsilon^{\alpha \beta z}
, \\
\mathrm{tr}\! \left[ \tau^{\alpha}_{\bk s} \mathcal{D}^{\beta \delta}_{i j, \bk s} \tau^{\delta}_{s s} \right]
	& = 2 \sin \vartheta_{\bk} ( \bm{P}_{i j} \cdot \hat{e}_{\varphi_{\bk}} ) \delta^{\alpha \beta}
		+ 2 s \sin \vartheta_{\bk} ( \bm{P}_{i j} \times \hat{e}_{\varphi_{\bk}} )^z \epsilon^{\alpha \beta z}
.\end{align}
For $\mathscr{Y}_{i}^{(1) \alpha}$, integrating by parts, we find
\begin{align}
\mathscr{Y}_{i}^{\alpha}
	& = e_{i j} A_{\mathrm{r}, j}^{\alpha} (\bq)
		+ f_{i j} \epsilon^{\alpha \beta z} A_{\mathrm{r}, j}^{\beta} (\bq)
,\end{align}
where
\begin{align}
e_{i j}
	& = \frac{e \hbar}{V} \sum_{\bk}
		e_{\bk s, i j} Q_{\bk s} (\zi \omega_{\lambda})
, \\
f_{i j}
	& = \frac{e \hbar}{V} \sum_{\bk}
		f_{\bk s, i j} Q_{\bk s} (\zi \omega_{\lambda})
\end{align}
with
\begin{align}
Q_{\bk s} (\zi \omega_{\lambda})
	& = \frac{1}{\beta} \sum_n g^{+}_{\bk,s} g_{\bk, s}
	\to - \zi \omega \tau \delta (\mu - \epsilon_{\bk s})
\end{align}
and
\begin{align}
e_{\bk, s, i j}
	& = \frac{2 s \cos \vartheta_{\bk}}{\hbar} \left( \frac{s}{\hbar} \frac{\partial \epsilon_{\bk s}}{\partial k_j} \right)
		\frac{\partial \cos \vartheta_{\bk}}{\partial k_i}
		- \frac{s \mathcal{J}_{sd}}{\hbar^2}
		\frac{\partial}{\partial k_i} \sin \vartheta_{\bk} (\bm{A}^{\perp}_{\mathrm{k}, j} \times \hat{e}_{\varphi_{\bk}})^z		
\notag \\ & \hspace{1em}
	+ \frac{s \Delta_{\bk}}{\hbar^2} \bm{A}_{\mathrm{k}, i}^{\perp} \cdot \bm{A}_{\mathrm{k}, j}^{\perp}
		\sin^2 \vartheta_{\bk}
	+ \frac{1}{\hbar^2} \cos \vartheta_{\bk}
		\sin \vartheta_{\bk} ( \bm{P}_{i j} \cdot \hat{e}_{\varphi_{\bk}} )
, \\
f_{\bk s, i j}
	& = \frac{\mathcal{J}_{sd}}{\hbar^2} \frac{\partial}{\partial k_i} \left[
		 \sin \vartheta_{\bk} (\bm{A}^{\perp}_{\mathrm{k}, j} \cdot \hat{e}_{\varphi_{\bk}})
	\right]
	+ \frac{s}{\hbar^2} \cos \vartheta_{\bk}
		\sin \vartheta_{\bk} ( \bm{P}_{i j} \times \hat{e}_{\varphi_{\bk}} )^z
.\end{align}
By using Eqs.~(\ref{eq:A_x_e}), (\ref{eq:A_dot_e}),
\begin{align}
\sin \vartheta_{\bk} (\bm{P}_{ij} \cdot \hat{e}_{\varphi_{\bk}})
	& = \frac{\mathcal{J}_{sd}}{\Delta_{\bk}^2} \frac{\partial \Delta_{\bk}}{\partial k_i} \frac{\partial \Delta_{\bk}}{\partial k_j}
		- \frac{\mathcal{J}_{sd}}{\Delta_{\bk}} \frac{\partial^2 \Delta_{\bk}}{\partial k_i \partial k_j}
		+ \mathcal{J}_{sd} \left( \frac{\partial \bm{x}_{\bk}}{\partial k_i} \cdot \frac{\partial \bm{x}_{\bk}}{\partial k_j} \right)
,\end{align}
and
\begin{align}
\sin \vartheta_{\bk} (\bm{P}_{ij} \times \hat{e}_{\varphi_{\bk}})^z
	& = \hat{z} \cdot \left(
		\Delta_{\bk} \bm{x}_{\bk} \times \frac{\partial^2 \bm{x}_{\bk}}{\partial k_i \partial k_j}
		+ \frac{\partial \Delta_{\bk}}{\partial k_i} \bm{x}_{\bk} \times \frac{\partial \bm{x}_{\bk}}{\partial k_j}
		+ \frac{\partial \Delta_{\bk}}{\partial k_j} \bm{x}_{\bk} \times \frac{\partial \bm{x}_{\bk}}{\partial k_i}
	\right)
,\end{align}
we have
\begin{align}
e_{\bk s ,i j}
	& = - 2 \frac{\cos^2 \vartheta_{\bk}}{\Delta_{\bk}}
		\frac{1}{\hbar^2} \frac{\partial \epsilon_{\bk s}}{\partial k_j}
		\frac{\partial \Delta_{\bk}}{\partial k_i}
	- \frac{s}{\hbar^2} \cos^2 \vartheta_{\bk} \left(
		\frac{2}{\Delta_{\bk}} \frac{\partial \Delta_{\bk}}{\partial k_i} \frac{\partial \Delta_{\bk}}{\partial k_j}
		- \frac{\partial^2 \Delta_{\bk}}{\partial k_i \partial k_j}
	\right)
	+ \frac{\Delta_{\bk}}{\hbar^2} \left( s \sin^2 \vartheta_{\bk} + \cos^2 \vartheta_{\bk} \right)
		\frac{\partial \bm{x}_{\bk}}{\partial k_i} \cdot \frac{\partial \bm{x}_{\bk}}{\partial k_j}
, \\
f_{\bk s ,i j}
	& = - \frac{\mathcal{J}_{sd}}{\hbar^2} \hat{z} \cdot \frac{\partial}{\partial k_i} \left( \bm{x}_{\bk} \times \frac{\partial \bm{x}_{\bk}}{\partial k_j} \right)
		+ \frac{s}{\hbar^2} \cos \vartheta_{\bk} \hat{z} \cdot \left(
		\Delta_{\bk} \bm{x}_{\bk} \times \frac{\partial^2 \bm{x}_{\bk}}{\partial k_i \partial k_j}
		+ \frac{\partial \Delta_{\bk}}{\partial k_i} \bm{x}_{\bk} \times \frac{\partial \bm{x}_{\bk}}{\partial k_j}
		+ \frac{\partial \Delta_{\bk}}{\partial k_j} \bm{x}_{\bk} \times \frac{\partial \bm{x}_{\bk}}{\partial k_i}
	\right)
.\end{align}
Similar consideration as in $\Phi' (\zi \omega_{\lambda})$ [Eqs.~(\ref{eq:Phi'_with_SIS}) and (\ref{eq:Phi'_without_SIS})] leads to $f_{\bk s ,i j} = 0$ and
\begin{align}
Y^{\alpha}_j (\bq, \omega)
	& = - \zi \omega \frac{\mathcal{P}_n}{e \mathcal{J}_{sd}} \sigma_{\mathrm{c}} A_{\mathrm{r}, j}^{\alpha} (\bq)
,\end{align}
where
\begin{align}
\mathcal{P}_n
	& = \frac{e^2 \mathcal{J}_{sd}}{\sigma_{\mathrm{c}}} \frac{1}{V} \sum_{\bk, s} e_{\bk, s, ii} \tau \delta (\mu - \epsilon_{\bk s})
\notag \\
	& = \frac{e^2 \mathcal{J}_{sd}}{\sigma_{\mathrm{c}}} \frac{1}{V} \sum_{\bk, s} \left[
	\frac{\mathcal{J}_{sd}^2}{\Delta_{\bk}^2} \frac{s}{\hbar^2} \frac{\partial^2 \Delta_{\bk}}{\partial k_i \partial k_i}
	- 2 \frac{\mathcal{J}_{sd}^2}{\Delta_{\bk}^3}
		\frac{1}{\hbar^2} \frac{\partial T_{\bk}}{\partial k_i}
		\frac{\partial \Delta_{\bk}}{\partial k_i}
	+ \frac{1}{\hbar^2} \frac{ s | \eta_{\bk} |^2 + \mathcal{J}_{sd}^2 }{\Delta_{\bk}}
		\frac{\partial \bm{x}_{\bk}}{\partial k_i} \cdot \frac{\partial \bm{x}_{\bk}}{\partial k_i}
	\right] \tau \delta (\mu - \epsilon_{\bk s})
.\end{align}
For the simple case of $T_{\bk} = 0$ and with the site inversion symmetry, one find
\begin{align}
\mathcal{P}_n
	& \simeq \frac{\mathcal{J}_{sd}^3}{|\mu|^3} \left( \frac{\mathcal{J}_{sd}^2}{\mu^2 - \mathcal{J}_{sd}^2} - \sign{\mu} \right)
.\end{align}
We here note that $\mathcal{P}_n$ does not have the particle-hole symmetry.
We also note that $\mathcal{P}_n \to \infty$ when $|\mu| \to \mathcal{J}_{sd}$, but the spin torque is to be zero since $\sigma_{\mathrm{c}} \to 0$.

Hence, we obtain
\begin{align}
\langle s^{\alpha} (\bq, \omega) \rangle_{\mathrm{ne}}
	& = \frac{\mathcal{P}_n}{e \mathcal{J}_{sd}} j_{\mathrm{c}, j} A_{\mathrm{r}, j}^{\alpha} (\bq)
,\end{align}
which leads to
\begin{align}
\bm{\tau}_n
	& = \frac{1}{N} \left( \frac{\mathcal{P}_n}{e} \bm{j}_{\mathrm{c}} \cdot \bm{\nabla} \right) \bm{n}
.\end{align}

\section{\label{sec:nonadiabatic_torques}Nonadiabatic spin-transfer torques}
Finally, we derive and calculate the nonadiabatic spin-transfer torques.
In Eqs.~(\ref{eq:spin_na}) and (\ref{eq:Sigma_na}), we have the nonadiabatic components of $\bm{s}$ and $\bm{\Sigma}$.
Hence, the nonadiabatic spin torques in the rotated frame are given as
\begin{align}
(\mathcal{R}^{-1} \bm{\tau}_m^{\mathrm{na}})_{\bq}
	= \frac{J_{sd} N}{\hbar} \hat{z} \times \langle \bm{\Sigma}_{\mathrm{na}}^{\perp} (\bq, t) \rangle_{\mathrm{neq}}
, \qquad
(\mathcal{R}^{-1} \bm{\tau}_n^{\mathrm{na}})_{\bq}
	= \frac{J_{sd}}{\hbar} \hat{z} \times \langle \bm{s}_{\mathrm{na}}^{\perp} (\bq, t) \rangle_{\mathrm{neq}}
,\end{align}
so that we calculate the linear response of $\bm{\Sigma}^{\perp}_{\mathrm{na}}$ and $\bm{s}^{\perp}_{\mathrm{na}}$ to the vector potential as
\begin{align}
\langle \Sigma^{\alpha}_{\mathrm{na}} (\bq, \omega) \rangle_{\mathrm{neq}}
	& = X^{\mathrm{na}, \alpha}_{j} (\bq, \omega) A^{\mathrm{em}}_j (\omega)
, \\
\langle s^{\alpha}_{\mathrm{na}} (\bq) \rangle_{\mathrm{neq}}
	& = Y^{\mathrm{na}, \alpha}_{j} (\bq, \omega) A^{\mathrm{em}}_j (\omega)
,\end{align}
where $\alpha = x, y$, and the response coefficients are obtained from the corresponding Matsubara functions
\begin{align}
\mathscr{X}^{\mathrm{na}, \alpha}_j (\bq, \zi \omega_{\lambda})
	& = \frac{- e}{V} \int_0^{\beta} \dd{\tau} e^{\zi \omega_{\lambda} \tau} \left\langle
		\mathrm{T}_{\tau} \left\{ \Sigma_{\mathrm{na}}^{\alpha} (\bq, \tau) j_{j} (0, 0) \right\}
	\right\rangle
\label{eq:X_na_Matsubara}
, \\
\mathscr{Y}^{\mathrm{na}, \alpha}_j (\bq, \zi \omega_{\lambda})
	& = \frac{- e}{V} \int_0^{\beta} \dd{\tau} e^{\zi \omega_{\lambda} \tau} \left\langle
		\mathrm{T}_{\tau} \left\{ s_{\mathrm{na}}^{\alpha} (\bq, \tau) j_{j} (0, 0) \right\}
	\right\rangle
\end{align}
by taking the analytic continuation, $\zi \omega_{\lambda} \to \hbar \omega + \zi 0$.
We below evaluate $\mathscr{X}^{\mathrm{na}, \alpha}_j (\bq, \zi \omega_{\lambda})$ and $\mathscr{Y}^{\mathrm{na}, \alpha}_j (\bq, \zi \omega_{\lambda})$.

\begin{figure}[htbp]
\centering
\includegraphics[width=\linewidth]{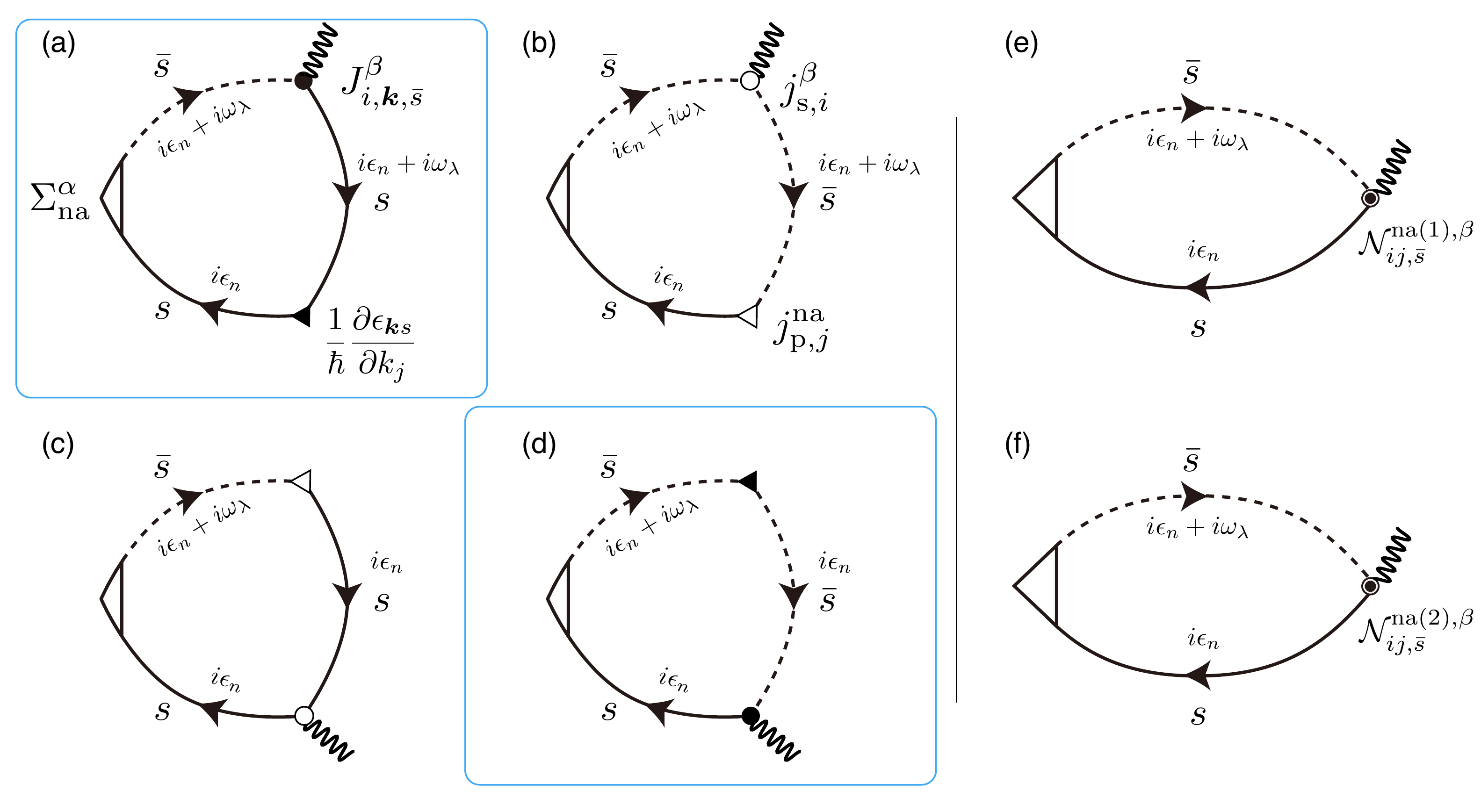}
\caption{\label{fig:diagrams}Feynman diagrams for (a)-(d) $\mathscr{X}^{\mathrm{na}, (1), \alpha}_j$, for (e) $\mathscr{X}^{\mathrm{na}, (2), \alpha}_j$ and for (f) $\mathscr{X}^{\mathrm{na}, (3), \alpha}_j$. The diagrams (a) and (d) are main contributions with respect to the lifetime of the bonding/antibonding state.}
\end{figure}
First, we calculate $\mathscr{X}^{\mathrm{na}, \alpha}_j (\bq, \zi \omega_{\lambda})$.
We divide $\mathscr{X}^{\mathrm{na}, \alpha}_j (\bq, \zi \omega_{\lambda})$ into three parts;
\begin{align}
\mathscr{X}^{\mathrm{na}, \alpha}_j (\bq, \zi \omega_{\lambda})
	& = \mathscr{X}^{\mathrm{na}, (1), \alpha}_j
	+ \mathscr{X}^{\mathrm{na}, (2), \alpha}_j
	+ \mathscr{X}^{\mathrm{na}, (3), \alpha}_j
,\end{align}
where
\begin{align}
\mathscr{X}^{\mathrm{na}, (1), \alpha}_j
	& = \frac{- e}{V} \int_0^{\beta} \dd{\tau} e^{\zi \omega_{\lambda} \tau} \langle \mathrm{T}_{\tau} \{ \Sigma^{\alpha}_{\mathrm{na}} (\bq, \tau) j_{\mathrm{p}, j} (0, 0) \} \rangle
, \\
\mathscr{X}^{\mathrm{na}, (2), \alpha}_j
	& = \frac{- e}{V} \int_0^{\beta} \dd{\tau} e^{\zi \omega_{\lambda} \tau} \langle \mathrm{T}_{\tau} \{ \Sigma^{\alpha}_{\mathrm{na}} (\bq, \tau) j_{\mathrm{A}, j}^{(1)} (0, 0) \} \rangle
, \\
\mathscr{X}^{\mathrm{na}, (3), \alpha}_j
	& = \frac{- e}{V} \int_0^{\beta} \dd{\tau} e^{\zi \omega_{\lambda} \tau} \langle \mathrm{T}_{\tau} \{ \Sigma^{\alpha}_{\mathrm{na}} (\bq, \tau) j_{\mathrm{A}, j}^{(2)} (0, 0) \} \rangle
.\end{align}
Rewriting them by means of the Green function, we have the Feynman diagrams shown in Fig.~\ref{fig:diagrams}, which read
\begin{align}
\mathscr{X}^{\mathrm{na}, (1), \alpha}_j
	& = \frac{\hbar}{2} \frac{e}{V} A^{\beta}_{\mathrm{r}, i} (\bq) \frac{1}{\beta} \sum_n \sum_{\bk, s}
		\tr[\mathcal{E}_{s}^{\alpha \gamma} \tau_{s \bar{s}}^{\gamma} J_{i, \bk \bar{s}}^{\beta}]
		\left(
			\frac{1}{\hbar} \frac{\partial \epsilon_{\bk \bar{s}}}{\partial k_j}
				g_{\bk, \bar{s}}^{+} g_{\bk, \bar{s}} g_{\bk, s} \Lambda_{s}
			+ \frac{1}{\hbar} \frac{\partial \epsilon_{\bk s}}{\partial k_j}
				g_{\bk, \bar{s}}^{+} g_{\bk, s}^{+} g_{\bk, s} \Lambda_{s}
	\right)
\notag \\ & \hspace{1em}
	+ \frac{\hbar}{2} \frac{e}{V} A^{\beta}_{\mathrm{r}, i} (\bq) \frac{1}{\beta} \sum_n \sum_{\bk, s}
		\tr \left[
			\mathcal{E}_s^{\alpha \gamma} \tau_{s \bar{s}}^{\gamma}
			\left(
				- \frac{s}{\hbar} \frac{\partial \epsilon_{\bk \bar{s}}}{\partial k_j} \tau_{\bk \bar{s}}^{\beta}
				+ \frac{\mathcal{J}_{sd}}{\hbar} \mathcal{C}^{\beta \delta}_{\bk \bar{s}, j} \tau_{\bar{s} \bar{s}}^{\delta}
			\right)
			\frac{- \Delta_{\bk}}{\hbar} (\hat{z} \times \bm{A}_{\mathrm{k}, j}^{\perp}) \cdot \bm{\tau}_{\bar{s} s}^{\perp}
		\right]
		(g^{+}_{\bk, \bar{s}})^2 g_{\bk, s} \Lambda_s
\notag \\ & \hspace{1em}
	+ \frac{\hbar}{2} \frac{e}{V} A^{\beta}_{\mathrm{r}, i} (\bq) \frac{1}{\beta} \sum_n \sum_{\bk, s}
		\tr \left[
			\mathcal{E}_s^{\alpha \gamma} \tau_{s \bar{s}}^{\gamma}
			\frac{- \Delta_{\bk}}{\hbar} (\hat{z} \times \bm{A}_{\mathrm{k}, j}^{\perp}) \cdot \bm{\tau}_{\bar{s} s}^{\perp}
			\left(
				\frac{s}{\hbar} \frac{\partial \epsilon_{\bk s}}{\partial k_j} \tau_{\bk s}^{\beta}
				+ \frac{\mathcal{J}_{sd}}{\hbar} \mathcal{C}^{\beta \delta}_{\bk s, j} \tau^{\delta}_{s s}			
			\right)
		\right]
		g^{+}_{\bk, \bar{s}} (g_{\bk, s})^2 \Lambda_s
, \\
\mathscr{X}^{\mathrm{na}, (2), \alpha}_j
	& = \frac{\hbar}{2} \frac{e}{V} A^{\beta}_{\mathrm{r}, i} (\bq) \frac{1}{\beta} \sum_n \sum_{\bk, s}
		\tr \left[
			\mathcal{E}_{s}^{\alpha \gamma} \tau_{s \bar{s}}^{\gamma} \mathcal{N}_{ij, \bar{s}}^{\mathrm{na} (1), \beta}
		\right]
		g_{\bk, \bar{s}}^{+} g_{\bk, s} \Lambda_{s}
, \\
\mathscr{X}^{\mathrm{na}, (3), \alpha}_j
	& = \frac{\hbar}{2} \frac{e}{V} A^{\beta}_{\mathrm{r}, i} (\bq) \frac{1}{\beta} \sum_n \sum_{\bk, s}
		\tr \left[
			\mathcal{E}_{s}^{\alpha \gamma} \tau_{s \bar{s}}^{\gamma} \mathcal{N}_{ij, \bar{s}}^{\mathrm{na} (2), \beta}
		\right]
		g_{\bk, \bar{s}}^{+} g_{\bk, s} \Lambda_{s}
,\end{align}
where
\begin{align}
\begin{pmatrix}
	- \tau_{s \bar{s}}^z
\\	\zi s \tau_{s \bar{s}}^0
\end{pmatrix}^{\alpha}
	& = \begin{pmatrix}
		0
	&	- 1
	\\	\zi s
	&	0
	\end{pmatrix}^{\alpha \beta}
	\begin{pmatrix}
		\tau_{s \bar{s}}^0
	\\	\tau_{s \bar{s}}^z
	\end{pmatrix}^{\beta}
	\equiv \mathcal{E}_{s}^{\alpha \beta} \tau_{s \bar{s}}^{\beta}
,\end{align}
$\bm{J}_{i, \bk, s} = (J_{i, \bk, s}^{x}, J_{i, \bk, s}^y, 0)$ is given by Eq.~(\ref{eq:J_i_k_s}), and $\Lambda_{s}$ is the vertex correction given by
\begin{align}
\Lambda_{s}
	& = 1 + n_{\mathrm{i}} u_{\mathrm{i}}^2 \sum_{\bk} g_{\bk, \bar{s}}^{+} g_{\bk, s} \Lambda_{s}
.\end{align}
Here, $\mathcal{N}_{ij, \bar{s}}^{\mathrm{na} (1), \beta}$ and $\mathcal{N}_{ij, \bar{s}}^{\mathrm{na} (2), \beta}$ are the nonadiabatic components of the anomalous current.
(We see below that $\mathscr{X}^{\mathrm{na}, (2), \alpha}_j$ and $\mathscr{X}^{\mathrm{na}, (3), \alpha}_j$ are negligible.)

The vertex correction $\Lambda_s$ is calculated as
\begin{align}
\Lambda_{s}
	& = \frac{1}{ 1 - n_{\mathrm{i}} u_{\mathrm{i}}^2 \sum_{\bk} g_{\bk, \bar{s}}^{+} g_{\bk, s} }
,\end{align}
and by using
\begin{align}
g^{+}_{\bk, \bar{s}} g_{\bk, s}
	& = \frac{ - g^{+}_{\bk, \bar{s}} + g_{\bk, s} }{ \zi \omega_{\lambda} - 2 s \Delta_{\bk} - \varSigma_{\bar{s}}^{+} + \varSigma_{s} }
	\simeq \frac{ - g^{+}_{\bk, \bar{s}} + g_{\bk, s} }{ \zi \omega_{\lambda} - 2 s \mathcal{J}_{sd} - \varSigma_{\bar{s}}^{+} + \varSigma_{s} }
,\end{align}
where $\varSigma_s^{+} = \varSigma_s (\zi \epsilon_n + \zi \omega_{\lambda}) = n_{\mathrm{i}} u_{\mathrm{i}}^2 \sum_{\bk} g^{+}_{\bk, s}$ and $\varSigma = \varSigma (\zi \epsilon_n) = n_{\mathrm{i}} u_{\mathrm{i}}^2 \sum_{\bk} g_{\bk, s}$ are the self energies, and we have approximate $\Delta_{\bk} \simeq \mathcal{J}_{sd}$, we have
\begin{align}
g^{+}_{\bk, \bar{s}} g_{\bk, s} \Lambda_{s}
	& = \frac{ - g^{+}_{\bk, \bar{s}} + g_{\bk, s} }{ \zi \omega_{\lambda} - 2 s \mathcal{J}_{sd} }
.\end{align}
From this, we find the main contribution is the first term of $\mathscr{X}^{\mathrm{na}, (1), \alpha}_j$, which is evaluated as
\begin{align}
\mathscr{X}^{\mathrm{na}, \alpha}_j (\bq, \zi \omega_{\lambda})
	& = \frac{\hbar}{2} \frac{e}{V} A^{\beta}_{\mathrm{r}, i} (\bq) \frac{1}{\beta} \sum_n \sum_{\bk, s}
		\tr[\mathcal{E}_{s}^{\alpha \gamma} \tau_{s \bar{s}}^{\gamma} J_{i, \bk \bar{s}}^{\beta}]
		\left(
			\frac{1}{\hbar} \frac{\partial \epsilon_{\bk s}}{\partial k_j}
				g_{\bk, s}^{+} g_{\bk, s}
			- \frac{1}{\hbar} \frac{\partial \epsilon_{\bk \bar{s}}}{\partial k_j}
				g^{+}_{\bk, \bar{s}} g_{\bk, \bar{s}}
	\right) \frac{ 1 }{ \zi \omega_{\lambda} - 2 s \mathcal{J}_{sd} }
\notag \\
	& = \frac{\hbar}{2} \frac{e}{V} A^{\beta}_{\mathrm{r}, i} (\bq) \frac{1}{\beta} \sum_n \sum_{\bk, s}
		\tr[\mathcal{E}_{s}^{\alpha \gamma} \tau_{s \bar{s}}^{\gamma} J_{i, \bk \bar{s}}^{\beta}]
		\frac{ 1 }{ \zi \omega_{\lambda} - 2 s \mathcal{J}_{sd} }
		\sum_{\sigma} \frac{1}{\hbar} \frac{\partial \epsilon_{\bk \sigma}}{\partial k_j} g_{\bk, \sigma}^{+} g_{\bk, \sigma}
.\end{align}
The trace $\tr [ \cdots ]$ is calculated as
\begin{align}
\tr[\mathcal{E}_{s}^{\alpha \gamma} \tau_{s \bar{s}}^{\gamma} J_{i, \bk \bar{s}}^{\beta}]
	& = \frac{2 \cos \vartheta_{\bk}}{\hbar} \frac{\partial \epsilon_{\bk \bar{s}}}{\partial k_i} \zi s \epsilon^{\alpha \beta z}
		- \frac{2 \Delta_{\bk}}{\hbar} \sin \vartheta_{\bk} (\bm{A}^{\perp}_{\mathrm{k}, i} \cdot \hat{e}_{\varphi_{\bk}}) \zi s \delta^{\alpha \beta}
		+ \frac{2 \Delta_{\bk}}{\hbar} \sin \vartheta_{\bk} (\bm{A}^{\perp}_{\mathrm{k}, i} \times \hat{e}_{\varphi_{\bk}})^z \zi \epsilon^{\alpha \beta z}
,\end{align}
which reads
\begin{align}
\tr[\mathcal{E}_{s}^{\alpha \gamma} \tau_{s \bar{s}}^{\gamma} J_{i, \bk \bar{s}}^{\beta}]
	& = \frac{2 \cos \vartheta_{\bk}}{\hbar} \frac{\partial \epsilon_{\bk \bar{s}}}{\partial k_i} \zi s \epsilon^{\alpha \beta z}
		- 2 \cos \vartheta_{\bk} \frac{\partial \Delta_{\bk}}{\partial k_i} \zi \epsilon^{\alpha \beta z}
	= \frac{2 \cos \vartheta_{\bk}}{\hbar} \frac{\partial T_{\bk}}{\partial k_i} \zi s \epsilon^{\alpha \beta z}
.\end{align}
(the term proportional to $\delta^{\alpha \beta}$ is zero from the similar consideration as in Eqs.~(\ref{eq:Phi'_with_SIS}) and (\ref{eq:Phi'_without_SIS}).)

Hence, we have
\begin{align}
\mathscr{X}^{\mathrm{na}, \alpha}_j (\bq, \zi \omega_{\lambda})
	& = \frac{e \hbar}{V} (\bm{A}^{\perp}_{\mathrm{r}, i} \times \hat{z})^{\alpha}
	\frac{1}{\beta} \sum_n \sum_{\bk, s}
		\frac{ 1 }{ \zi \omega_{\lambda} - 2 s \mathcal{J}_{sd} }
		\sum_{\sigma} \frac{1}{\hbar^2} \frac{\partial T_{\bk}}{\partial k_i} \frac{\partial \epsilon_{\bk \sigma}}{\partial k_j} g_{\bk, \sigma}^{+} g_{\bk, \sigma}
,\end{align}
and taking the analytic continuation, we finally obtain
\begin{align}
X^{\mathrm{na}, \alpha}_j (\bq, \omega)
	& = - \zi \omega \frac{2 \zi \hbar^2 \omega}{\hbar^2 \omega^2 - 4 \mathcal{J}_{sd}^2} \frac{\mathcal{P}_m^{\mathrm{na}}}{e} \sigma_{\mathrm{c}} (\bm{A}^{\perp}_{\mathrm{r}, j} \times \hat{z})^{\alpha}
,\end{align}
where
\begin{align}
\mathcal{P}_m^{\mathrm{na}}
	& = \frac{e^2}{\sigma_{\mathrm{c}}} \frac{1}{V} \sum_{\bk, s}
		\frac{\mathcal{J}_{sd}}{\Delta_{\bk}} \frac{1}{\hbar^2} \frac{\partial T_{\bk}}{\partial k_i} \frac{\partial \epsilon_{\bk \sigma}}{\partial k_i} \tau \delta (\mu - \epsilon_{\bk s})
.\end{align}
For the case of $T_{\bk} = 0$, we find $\mathcal{P}_m^{\mathrm{na}} = 0$.

From above, the nonadiabatic spin-transfer torque on the ferromagnetic vector is obtained as
\begin{align}
\bm{\tau}_m^{\mathrm{na}} (\br, t)
	& = \frac{\zi \omega \tau_{sd}}{1 - \omega^2 \tau_{sd}^2} \bm{n} \times \left( \frac{\mathcal{P}^{\mathrm{na}}_m}{e} \bm{j}_{\mathrm{c}} \cdot \bm{\nabla} \right) \bm{n}
	\simeq \zi \omega \tau_{sd} \bm{n} \times \left( \frac{\mathcal{P}^{\mathrm{na}}_m}{e} \bm{j}_{\mathrm{c}} \cdot \bm{\nabla} \right) \bm{n}
	\qquad (\omega \tau_{sd} \ll 1)
.\end{align}

The calculation of $\mathscr{Y}^{\mathrm{na}, \alpha}_j (\bq, \zi \omega_{\lambda})$ can be done in the similar way of $\mathscr{X}^{\mathrm{na}, \alpha}_j (\bq, \zi \omega_{\lambda})$, and we have
\begin{align}
\mathscr{Y}^{\mathrm{na}, \alpha}_j (\bq, \zi \omega_{\lambda})
	& \simeq \frac{\hbar}{2} \frac{e}{V} A^{\beta}_{\mathrm{r}, i} (\bq) \frac{1}{\beta} \sum_n \sum_{\bk, s}
		\tr[\mathcal{F}_{s}^{\alpha \gamma} \tau_{s \bar{s}}^{\gamma} J_{i, \bk \bar{s}}^{\beta}]
		\left(
			\frac{1}{\hbar} \frac{\partial \epsilon_{\bk s}}{\partial k_j}
				g_{\bk, s}^{+} g_{\bk, s}
			- \frac{1}{\hbar} \frac{\partial \epsilon_{\bk \bar{s}}}{\partial k_j}
				g^{+}_{\bk, \bar{s}} g_{\bk, \bar{s}}
	\right) \frac{ 1 }{ \zi \omega_{\lambda} - 2 s \mathcal{J}_{sd} }
\notag \\
	& = \frac{\hbar}{2} \frac{e}{V} A^{\beta}_{\mathrm{r}, i} (\bq) \frac{1}{\beta} \sum_n \sum_{\bk, s}
		\tr[\mathcal{F}_{s}^{\alpha \gamma} \tau_{s \bar{s}}^{\gamma} J_{i, \bk \bar{s}}^{\beta}]
		\frac{ 1 }{ \zi \omega_{\lambda} - 2 s \mathcal{J}_{sd} }
		\sum_{\sigma} \frac{1}{\hbar} \frac{\partial \epsilon_{\bk \sigma}}{\partial k_j} g_{\bk, \sigma}^{+} g_{\bk, \sigma}
,\end{align}
where $\mathcal{F}_s^{\alpha \gamma}$ is defined by
\begin{align}
\begin{pmatrix}
	- \cos \vartheta_{\bk} \tau_{s \bar{s}}^0
\\	\zi s \cos \vartheta_{\bk} \tau_{s \bar{s}}^z
\end{pmatrix}^{\alpha}
	& = \begin{pmatrix}
		- \cos \vartheta_{\bk}
	&	0
	\\	0
	&	\zi s \cos \vartheta_{\bk}
	\end{pmatrix}^{\alpha \beta}
	\begin{pmatrix}
		\tau_{s \bar{s}}^0
	\\	\tau_{s \bar{s}}^z
	\end{pmatrix}^{\beta}
	\equiv \mathcal{F}_s^{\alpha \beta} \tau_{s \bar{s}}^{\beta}
.\end{align}
Here, the trace is evaluated as
\begin{align}
\tr[\mathcal{F}_{s}^{\alpha \gamma} \tau_{s \bar{s}}^{\gamma} J_{i, \bk \bar{s}}^{\beta}]
	& = \frac{2 \cos^2 \vartheta_{\bk}}{\hbar} \frac{\partial \epsilon_{\bk \bar{s}}}{\partial k_i} \delta^{\alpha \beta}
		- \frac{2 \Delta_{\bk}}{\hbar} \sin \vartheta_{\bk} \cos \vartheta_{\bk} (\bm{A}^{\perp}_{\mathrm{k}, i} \cdot \hat{e}_{\varphi_{\bk}}) \epsilon^{\alpha \beta z}
		+ \frac{2 \Delta_{\bk}}{\hbar} \sin \vartheta_{\bk} \cos \vartheta_{\bk} (\bm{A}^{\perp}_{\mathrm{k}, i} \times \hat{e}_{\varphi_{\bk}})^z \delta^{\alpha \beta}
\notag \\
	& = \frac{2 \cos^2 \vartheta_{\bk}}{\hbar} \frac{\partial T_{\bk}}{\partial k_i} \delta^{\alpha \beta}
.\end{align}
(the term proportional to $\epsilon^{\alpha \beta z}$ is zero from the similar consideration as in Eqs.~(\ref{eq:Phi'_with_SIS}) and (\ref{eq:Phi'_without_SIS}).)

Finally, we have
\begin{align}
Y^{\mathrm{na}, \alpha}_j (\bq, \omega)
	& = - \zi \omega \frac{4 \hbar \mathcal{J}_{sd} }{\hbar^2 \omega^2 - 4 \mathcal{J}_{sd}^2} \frac{\mathcal{P}_n^{\mathrm{na}}}{e} \sigma_{\mathrm{c}} \bm{A}_{\mathrm{r}, j}^{\alpha}
,\end{align}
which leads to the nonadiabatic spin-transfer torque on the Neel vector,
\begin{align}
\bm{\tau}_n (\br, t)
	& = \frac{1}{N} \frac{1}{1 - \omega^2 \tau_{sd}^2} \left( \frac{\mathcal{P}_{n}^{\mathrm{na}}}{e} \bm{j}_{\mathrm{c}} \cdot \bm{\nabla} \right) \bm{n}
	\simeq - \frac{1}{N} \left( \frac{\mathcal{P}_{n}^{\mathrm{na}}}{e} \bm{j}_{\mathrm{c}} \cdot \bm{\nabla} \right) \bm{n}
	\qquad (\omega \tau_{sd} \ll 1)
.\end{align}
Here,
\begin{align}
\mathcal{P}_n^{\mathrm{na}}
	& = \frac{e^2}{\sigma_{\mathrm{c}}} \frac{1}{V} \sum_{\bk, s} \left( \frac{\mathcal{J}_{sd}}{\Delta_{\bk}} \right)^2
		\frac{1}{\hbar^2} \frac{\partial T_{\bk}}{\partial k_i} \frac{\partial \epsilon_{\bk \sigma}}{\partial k_i} \tau \delta (\mu - \epsilon_{\bk s})
,\end{align}
and For the case of $T_{\bk} = 0$, we have $\mathcal{P}_n^{\mathrm{na}} = 0$, and $\Delta_{\bk} \simeq \mathcal{J}_{sd}$, we find
\begin{align}
\mathcal{P}_n^{\mathrm{na}}
	& = \mathcal{P}_m^{\mathrm{na}}
	\simeq 1
,\end{align}
since $\partial T_{\bk} / \partial k_i \simeq \partial \epsilon_{\bk s} / \partial k_i$.

